\documentclass[utf8]{frontiersFPHY} % for Physics and Applied Mathematics and Statistics articles

\setcitestyle{square} % for Physics and Applied Mathematics and Statistics articles
\usepackage{url,hyperref,lineno,microtype,subcaption}
\usepackage[onehalfspacing]{setspace}
\usepackage{amsmath,amsfonts,amsthm,bm}
\usepackage[utf8]{inputenc}
\usepackage{bbold}
\usepackage{graphicx,amssymb,bm,bbm}
\usepackage{multirow,booktabs,dcolumn}
\usepackage{appendix}
\usepackage{slashed}
\def\keyFont{\fontsize{8}{11}\helveticabold }
\def\firstAuthorLast{Piarulli {et~al.}} %use et al only if is more than 1 author
\def\Authors{Maria Piarulli\,$^{1,*}$ and Ingo Tews\,$^{2}$}
\def\Address{$^{1}$ Physics Department, Washington University, St Louis, MO 63130, USA \\
$^{2}$ Theoretical Division, Los Alamos National Laboratory, Los Alamos, NM 87545, USA}
\def\corrAuthor{Maria Piarulli}
\def\corrEmail{mpiarulli@physics.wustl.edu}

\begin{document}
\onecolumn
\firstpage{1}

\title[Local chiral interactions]{Local nucleon-nucleon and three-nucleon interactions within chiral effective field theory} 

\author[\firstAuthorLast ]{\Authors} %This field will be automatically populated
\address{} %This field will be automatically populated
\correspondance{} %This field will be automatically populated

\extraAuth{}% If there are more than 1 corresponding author, comment this line and uncomment the next one.
%\extraAuth{corresponding Author2 \\ Laboratory X2, Institute X2, Department X2, Organization X2, Street X2, City X2 , State XX2 (only USA, Canada and Australia), Zip Code2, X2 Country X2, email2@uni2.edu}

\maketitle

\begin{abstract}
To obtain an understanding of the structure and reactions of nuclear systems from first principles has been a long-standing goal of nuclear physics. 
In this respect, few- and many-body systems provide a unique laboratory for studying nuclear interactions. 
During the past decades, the development of accurate representations of the nuclear force has undergone substantial progress. 
Particular emphasis has been devoted to chiral effective field theory (EFT), a low-energy effective representation of quantum chromodynamics (QCD).
Within chiral EFT, many studies have been carried out dealing with the construction of both the nucleon-nucleon ($N\!N$) and three-nucleon ($3N$) interactions. 
The aim of the present article is to give a detailed overview of the chiral interaction models that are local in configuration space, and show recent results for nuclear systems obtained by employing these local chiral forces.

\tiny
 \keyFont{ \section{Keywords:} nuclear interactions, chiral effective field theory, local interactions, three-body forces, ab-initio calculations} %All article types: you may provide up to 8 keywords; at least 5 are mandatory.
\end{abstract}

\section{Introduction}

The last few decades have marked the emergence of the {\it basic model} of nuclear theory in which nuclear systems -- particularly atomic nuclei and infinite nucleonic matter -- can be described as a collection of point-like particles, the nucleons, interacting with each other in terms of two- and many-body 
%(primarily, two- and three-body) 
effective interactions, and with external electroweak probes via effective current operators. 
This approach, in conjunction with a computational method of choice to solve the many-body Schr\"{o}dinger equation, can then be used to study the structure and dynamics of nuclear systems in a fully microscopic way,
%where the nucleons emerge as effective degrees of freedom, at sufficiently low-energy. 
%These calculations are commonly referred to as {\it ab-initio}. 
which is commonly referred to as {\it ab-initio} calculations.
Examples of such calculations are based on the no-core shell model (NCSM)~\cite{Barrett:2013nh,Jurgenson:2013yya}, the coupled cluster (CC)~\cite{Hagen:2013yba,Hagen:2013nca} or
hyperspherical harmonics (HH)~\cite{Kievsky:2008es} expansions,
similarity renormalization group (SRG) approaches~\cite{BOGNER201094,Hergert:2012nb},
self-consistent Green's function techniques~\cite{DICKHOFF2004377,Soma:2012zd}, quantum Monte Carlo (QMC) methods~\cite{Carlson:2014vla}, and nuclear lattice effective field theory (NLEFT)~\cite{Lahde:2013uqa}. 
Although significant progress has been made in recent years, these ab-initio techniques remain challenging and their domain of applicability is, at present, limited to provide quantitative description of light and medium-mass nuclei~\cite{Barrett:2013nh,Hagen:2013nca,Hergert:2012nb,Soma:2012zd,DICKHOFF2004377,Carlson:2014vla,Hagen:2015yea} and their reactions~\cite{Lovato:2014eva,Lovato:2019fiw,Hupin:2014iqa,Elhatisari:2015iga}. A special but related challenge is the development of microscopic models that include continuum couplings which are mandatory to describe, for instance, weakly bound nuclear systems~\cite{Fossez:2017wpa,Fossez:2018gae}.

One might argue that nucleons are not the fundamental building blocks of the nuclear systems at hand, and that one should instead start from Quantum Chromodynamics (QCD). QCD provides the theoretical framework to describe strong
interactions which governs the dynamics and properties of quarks and gluons. However, while strong interactions are weak and perturbative at high energies, i.e., short distances (asymptotic freedom), quarks are strongly interacting at low energies or long distances, of relevance for nuclear physics, which makes a nonperturbative treatment necessary. In addition, at these energies quarks are confined into colorless objects called hadrons (baryons, consisting of three quarks, e.g., the nucleon, and mesons consisting of a quark and an anti-quark, e.g. the pion).
Hence, while QCD is responsible for the complex inter-nucleon forces in nuclear systems, which can be thought of as residual interactions among quarks, a description in terms of nucleon degrees of freedom is particularly valid at sufficiently low energies. 

How the interactions among nucleons emerge from the fundamental theory, QCD, has kept nuclear theorists occupied for many decades. 
Since QCD is nonperturbative at low energies of interest in nuclear systems, one may try to solve QCD with brute computing power on a discretized Euclidean space-time lattice (known as lattice QCD)
However, in spite of many advances~\cite{Inoue:2013nfe,Beane:2014ora,Orginos:2015aya,Savage:2016kon}, lattice QCD calculations are still limited to small nucleon numbers and/or large pion masses, and hence, at the present time, can only be used to address a limited set of representative key-issues.

As a consequence, most theoretical studies of nuclear systems have to resort to using the basic model of nuclear theory, i.e., assuming pointlike nucleons to be the relevant degrees of freedom instead of quarks. 
In this review, we will briefly introduce this basic model and discuss the current state-of-the-art for nuclear interactions, chiral effective field theory (EFT). We will then focus on a particular subclass of chiral EFT interactions, local chiral EFT interactions, intended for the use in QMC methods. 

The review is structured as follows. In Section~\ref{sec:Hamiltonians}, we discuss the general features of nuclear interactions starting with the phenomenological ones and moving to those obtained in chiral EFT. In Section~\ref{sec:local}, we provide many details about the theoretical derivation of local interactions in both delta-full and delta-less chiral EFT, i.e., when explicitly including the delta resonance or not.
 In Section~\ref{sec:reg_artifacts}, we briefly discuss finite cutoff and regulator artifacts that can appear in calculations with local interactions. Finally, in Section~\ref{sec:results}, we report selected results for light and medium-mass nuclei and the equation of state of pure neutron matter using QMC methods.

\section{Nuclear Hamiltonians}
\label{sec:Hamiltonians}

The basic model of nuclear theory assumes that a nuclear system can be described by a non-relativistic Hamiltonian that contains interactions among nucleons, i.e., protons and neutrons. 
The individual nucleons mostly interact via two-body ($N\!N$) interactions. However, nucleons can also interact via three-body ($3N$) and higher many-body interactions. 
The way these many-body interactions appear is twofold.
First, nucleons are compound particles and, hence, treating them as point-like particles induces effective many-body interactions even if only two-quark interactions were to be considered. This is similar to describing tides on Earth, where the three-body system given by Earth, Moon, and Sun is relevant, even though gravity is only a two-body force. 
Second, since quarks themselves can have multi-quark interactions, this immediately leads to the appearance of 'true' $3N$ forces among nucleons, where, for example, single quarks in each of the 3 nucleons interact with each other.

The resulting Hamiltonian can then be written as a sum of the non-relativistic one-body kinetic energy  ($\mathcal{T}_i$), $N\!N$ interactions between particle $i$ and $j$ ($V_{ij}$), $3N$ interactions between particle $i$, $j$, and $k$ ($V_{ijk}$), and additional many-body interactions, and provides a good approximation for interacting nucleons in a given nuclear system:
\begin{equation}
    \mathcal{H}=\sum_i\mathcal{T}_i+\sum_{i<j}V_{ij}+\sum_{i<j<k}V_{ijk}+ \cdots\,.
\end{equation}
There are indications that four-body interactions may contribute at the level of only $\sim\,$100 keV in $^4$He~\cite{Rozpedzik:2006yi} or pure neutron matter~\cite{Kruger:2013kua}, and therefore are negligible compared to $N\!N$ and $3N$ interactions. Hence, current formulations of the basic model do not typically include them (see, for example, Ref.~\cite{Carlson:2014vla}). 

In order to derive two- and three-body nuclear forces, one has to take into account some general considerations, specify the theoretical framework in which such interactions are formulated, and the experimental inputs necessary to determine possible unknown parameters of the theory.

\subsection{General considerations for nuclear interactions}\label{sec:GenCons}

To accurately describe nuclear systems that are governed by QCD, nuclear interactions need to obey all the relevant symmetries of QCD. 
Hence, nuclear potentials need to have the following properties (we will focus on $N\!N$ forces here, but the statements remain true for all parts of the interaction):
\begin{itemize}
    \item $V$ is hermitian, because the Hamiltonian is hermitian,
    \item $V$ is symmetric under the permutation of identical particles, i.e., $V_{ij}=V_{ji}$,
    \item $V$ is translationally and rotationally invariant,
    \item $V$ is invariant under translations in time, i.e., time-independent,
    \item $V$ is Lorentz invariant (for nonrelativistic interactions this reduces to Galilean invariance),
    \item $V$ is invariant under parity transformations and time reversal,
    \item $V$ has to conserve baryon and lepton number,
    \item $V$ has to be approximately isospin symmetric and charge independent,
    \item and $V$ has to include the properties of spontaneously and explicitly broken chiral symmetry.
\end{itemize}

Chiral symmetry is a symmetry of the QCD Lagrangian with massless quarks under independent rotations of left- and right-handed quarks. Considering only u and d quarks, this symmetry can be written as SU$(2)_{L}\times$SU$(2)_{R}$. This expression contains two symmetries: the first (vector) one represents isospin symmetry, i.e., symmetry under the exchange of u and d quarks, and the second (axial) one is the so-called chiral symmetry. These two symmetries imply degenerate fermions under isospin and spin-parity transformations. While isospin symmetry is approximately fulfilled in nature, i.e., the neutron and proton have similar masses, nucleons with spin $1/2^+$ and $1/2^-$ have very different masses (940 MeV vs 1535 MeV). This signals that chiral symmetry is broken in nature. 

In fact, chiral symmetry is broken twofold. First, it is broken spontaneously, leading to the formation of Goldstone bosons, that can be identified with the pions. Second, chiral symmetry is also explicitly broken by the finite quark masses, which leads to the pion being pseudo-Goldstone bosons with finite but small mass. 
In contrast, isospin symmetry remains a good symmetry, because the ratio $(m_d-m_u)/\Lambda_{\textrm{QCD}}$ is very small, where $m_u \simeq 2.4$ MeV and $m_d \simeq 4.8$ MeV.

These symmetries only allow certain operator structures for nuclear interactions. Galilean invariance, for instance, implies that nuclear interactions depend only on relative momenta between two nucleons, $\textbf{p}=\textbf{p}_i-\textbf{p}_j$, while symmetry under parity transformations implies that nuclear interactions cannot be linear in $\textbf{p}$, and charge independence requires that the nuclear interactions can be written as
\begin{equation}
    V=V_{\mathbb{1}}\cdot \mathbb{1} +V_{\tau}\, {\bm \tau}_i\cdot {\bm \tau}_j\,,
\end{equation}
and so on. In addition, the spin dependencies are included through operators like ${1, {\bm \sigma}_i \cdot {\bm \sigma}_j}$, spin-orbit interactions given by ${\bf L}\cdot {\bf S}$ with ${\bf L}={\bf r} \times {\bf p}$, where ${\bf r}={\bf r}_i-{\bf r}_j$, or tensor interactions with the tensor operator $S_{ij}(\textbf{r})={\bm \sigma}_i\cdot \hat{\bf r}\, {\bm \sigma}_j\cdot \hat{\textbf{r}}-{\bm \sigma}_i\cdot {\bm \sigma}_j$. As a consequence, interactions typically have a spin-isospin operator structure given by 
\begin{equation}
    \mathcal{O}_V=\lbrace \mathbb{1}, {\bm \sigma}_i\cdot{\bm \sigma}_j, \textbf{L}\cdot \textbf{S}, S_{ij}\rbrace \times \lbrace\mathbb{1}, {\bm \tau}_i\cdot {\bm \tau}_j \rbrace\,,
\end{equation}
where the individual operators carry momentum-dependent functions consistent with all required symmetries.

\subsection{Phenomenological interactions}

Historically, $NN$ interactions were derived using phenomenological insight.
They were characterized by a long-range component characterizing the interaction for inter-nucleon separations $r \gtrsim 1/m_{\pi}$, due to one-pion exchange (OPE)~\cite{Yukawa:1935xg}, and intermediate- and short-range components describing the interactions at $1 \,\,{\rm fm}  \lesssim r  \lesssim 2$ fm and $ r \lesssim 1$ fm, respectively.  The intermediate- and short-range components were included to simulate intermediate-range attraction as well as short-range repulsion. 

Up to the mid-1990's, nuclear interactions were based almost exclusively on meson-exchange phenomenology. 
Interactions of the mid-1990's~\cite{Stoks:1994wp,Wiringa:1994wb,Machleidt:2000ge} were constrained by fitting nucleon-nucleon ($N\!N$) elastic scattering data up to laboratory energies of 350 MeV, with
$\chi^2$/datum $\simeq 1$ relative to the database available at the time~\cite{Stoks:1993tb}. Two well-known and still widely used examples in this class are the Argonne $v_{18}$ (AV18)~\cite{Wiringa:1994wb} and CD-Bonn~\cite{Machleidt:2000ge} interactions. These are so-called {\sl phenomenological} interactions. 

Already in the 1980's, accurate three-body calculations showed that contemporary $N\!N$ interactions alone did not provide sufficient binding to reproduce experimental numbers for nuclei with nucleon number $A=3$, $^3$H and $^3$He~\cite{Friar:1984ic}.
This realization was later on extended to the spectra (ground and low-lying excited states) of light p-shell nuclei, for instance, in calculations based on quantum Monte Carlo (QMC) methods~\cite{Pudliner:1997ck} and in no-core shell-model (NCSM) studies~\cite{Navratil:2000gs}.
Consequently, the basic model with only $N\!N$ interactions fit to scattering data, without the inclusion of a three-nucleon ($3N$) interaction, was found to be  unsatisfactory. 
However, because of the composite nature of the nucleon and, in particular, the dominant role of the $\Delta$ resonance, a spin-3/2, isospin-3/2 nucleon resonance, in pion-nucleon scattering, many-body interactions arise quite naturally in meson-exchange phenomenology.  

For example, the Illinois $3N$ interaction~\cite{Pieper:2001ap} consists of a dominant two-pion exchange (TPE) -- the Fujita-Miyazawa interaction~\cite{Fujita:1957zz} -- and smaller multi-pion exchange components resulting from the excitation of intermediate $\Delta$'s. 
The most recent version, Illinois-7 (IL7)~\cite{doi:10.1063/1.2932280}, also contains phenomenological isospin-dependent central terms. 
The parameters characterizing this $3N$ potential have been determined by fitting the low-lying spectra of nuclei in the mass range $A\,$=$\,$3--10. 
The resulting AV18+IL7 Hamiltonian, generally utilized with QMC methods, then leads to predictions of 100 ground- and excited-state energies up to $A\,$=$\,$12, including the $^{12}$C ground- and Hoyle-state energies, in good agreement with the corresponding experimental values~\cite{Carlson:2014vla}. 
However, when used to compute the neutron-star equation of state, these interactions do not provide sufficient repulsion to guarantee
the stability of the observed neutron stars with masses larger than two solar masses against gravitational collapse~\cite{Maris:2013rgq}. 
Thus, in the context of the phenomenological nuclear interactions, we do not have a Hamiltonian that can predict the properties of all nuclear systems, from $N\!N$ scattering to dense nuclear and neutron matter.

Furthermore, high-precision phenomenological potentials suffer from several limitations, most notably the missing connection with the low-energy QCD and, hence, the absence of a guiding principle for the construction of interactions. 
As a consequence, phenomenological interactions do not provide rigorous schemes to consistently derive two- and three-body forces and compatible electroweak currents. 
In addition, there is no clear way to properly assess the theoretical uncertainty associated with the nuclear potentials and currents.

\subsection{Chiral effective field theory}

These drawbacks were addressed when a new phase in the evolution of the basic model began in the early 1990's with the emergence of chiral effective field theory (EFT)~\cite{Weinberg:1990rz,Weinberg:1991um,Weinberg:1992yk}. 

Chiral EFT is a low-energy effective theory of QCD based on the choice of baryons as effective degrees of freedom: in chiral EFT one chooses pions and nucleons. 
At typical momenta in nuclei, $p\sim m_{\pi}\sim \mathcal{O}(100\, \mathrm{MeV})$, this choice is accurate, because shorter-range structures, e.g., the quark substructure, or heavier meson exchanges, e.g., exchanges of the $\rho$-meson, are not resolved, and can be absorbed in short-range nucleon contact interactions. 
This {\it separation of scales} between typical momenta $p$ and scales of the same order, i.e., the pion mass $m_{\pi}\sim 140$ MeV, and larger scales, e.g., the mass of the $\rho$, $m_{\rho}\sim 770$ MeV, can then be used to systematically derive an effective and most general scheme accommodating all possible interactions among the relevant degrees of freedom consistent with the symmetries of QCD. 
In some modern approaches, the choice of degrees of freedom also includes the $\Delta$ isobar (delta-full chiral EFT), because the $\Delta$-nucleon mass splitting is only $300$ MeV$\sim 2 m_{\pi}$.

The starting point in chiral EFT is the most general Lagrangian in terms of the chosen degrees of freedom, which contains all allowed interaction mechanisms in accordance with the considerations in Sec.~\ref{sec:GenCons}.
As a consequence, this Lagrangian contains an infinite number of terms and needs to be truncated using a given power-counting scheme. 
Most chiral interactions used in nuclear structure calculations are based on Weinberg power counting, which itself is based on naive dimensional analysis of interaction contributions. 
Within Weinberg power counting, the interactions are expanded in powers of the typical momentum $p$ over the breakdown scale $\Lambda_b$, $Q=p/\Lambda_b$, where the breakdown scale denotes momenta at which the short distance structure becomes important and cannot be neglected and absorbed into contact interactions anymore (see Refs.~\cite{Epelbaum:2008ga,Machleidt:2011zz,Machleidt:2016rvv,Machleidt:2017vls} for recent review articles). 
It is worthwhile mentioning that alternative power-counting schemes have been also suggested as in Refs.~\cite{Kaplan:1998tg,Kaplan:1998we,Nogga:2005hy,PavonValderrama:2005wv,Long:2011xw,vanKolck:1994yi}.

This expansion defines an order by order scheme, defined by the power $\nu$ of the expansion scale $Q$ in each interaction contribution: leading order (LO) for $\nu=0$, next-to-leading order (NLO) for $\nu=2$, next-to-next-to-leading order (N$^2$LO) for $\nu=3$ and so on. Similarly as for nuclear interactions, such a scheme can also be developed for electroweak currents. 
Therefore, chiral EFT provides a rigorous scheme to systematically construct many-body forces and consistent electroweak currents, and tools to estimate their uncertainties~\cite{Furnstahl:2014xsa,Epelbaum:2014efa,Furnstahl:2015rha,Wesolowski:2015fqa,Melendez:2017phj,Wesolowski:2018lzj}. 
From this perspective, it can be justifiably argued that chiral EFT has put the basic model on a more fundamental basis, by providing a link between QCD
with all its symmetries, and the strong and electroweak interactions in nuclei.

\begin{figure}[ht!]
\begin{center}
\includegraphics[width=16cm]{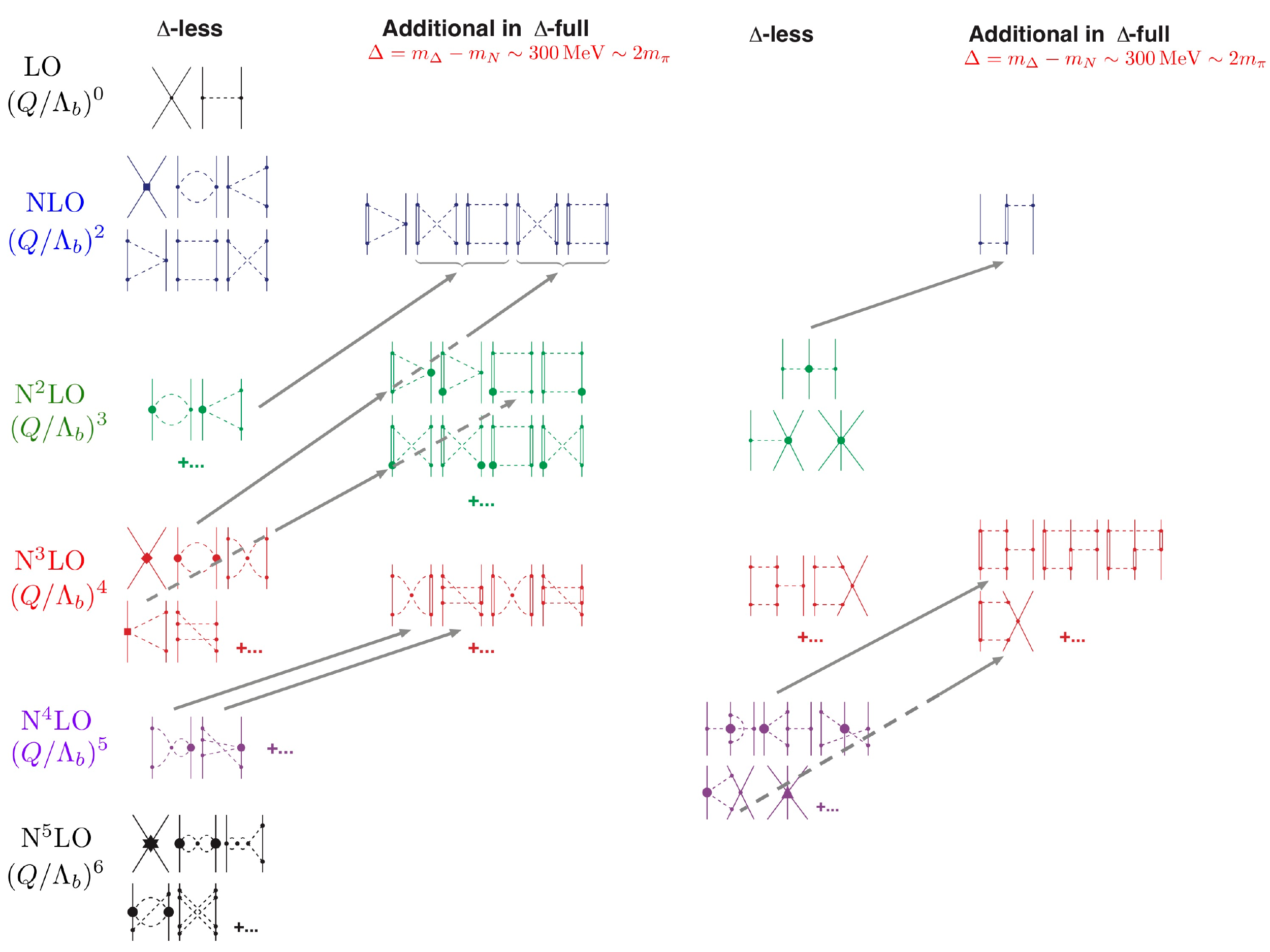}
\end{center}
\caption{Chiral contributions to the $N\!N$ and $3N$ interactions in the delta-less and delta-full chiral EFT based on Weinberg power counting. Solid lines represent nucleons, dashed lines represent pions, and double lines represent the $\Delta$ isobar. Grey arrows indicate the shift of individual contributions within the two power-counting schemes when explicit $\Delta$'s are accounted for. Figure adapted from Refs.~\cite{Machleidt:2011zz,Machleidt:2016rvv} under the Creative Commons CCBY license.}\label{fig:PowerCounting}
\end{figure}

Figure~\ref{fig:PowerCounting} shows the state of the art of chiral contributions to the $N\!N$ and $3N$ interactions in the delta-less and delta-full chiral EFT. Higher many-body forces, such as four-nucleon ($4N$) or five-nucleon ($5N$) interactions,  can naturally also be derived within this framework~\cite{Machleidt:2016rvv}, but they will not be discussed here. 
Nuclear forces in chiral EFT are separated into pion-exchange contributions and contact terms. 
Pion-exchange contributions represent the long- and intermediate-range parts of nuclear interactions and contain all chiral physics. 
Contact terms, on the other hand, encode the unresolved short-range physics and their strength is specified by unknown low-energy constants (LECs), that need to be adjusted to experimental data.

At LO, besides the already mentioned OPE potential, there are two $N\!N$ contact terms with no momentum dependence that contribute only to the $S$-wave. 
They are identified by the four-nucleon-leg diagram with a momentum-independent vertex denoted by a small dot in the first row of Fig.~\ref{fig:PowerCounting}. 
The interaction at LO is a very simple approximation, but already takes into account some of the important features of the $N\!N$ force. 
For instance, the OPE generates the tensor component of the nuclear force known to be crucial to properly describe the two-nucleon bound state (deuteron). 

The leading $N\!N$ two-pion-exchange (TPE) contributions appear at NLO. 
Diagrams involving virtual excitations of the $\Delta$-isobars~\cite{Ordonez:1993tn,Ordonez:1992xp,Ordonez:1995rz,Kaiser:1998wa} also appear at NLO in the delta-full chiral-EFT approach.
Most importantly, seven new momentum-dependent contact terms can be constructed at this order, which are denoted by the four-nucleon-leg graph with a solid square in the second row of Fig.~\ref{fig:PowerCounting}. These additional contact terms are important to correctly describe $N\!N$ scattering in the $S$- and $P$-waves. More details about these contributions are presented in the next sections. 
Another important contribution at NLO is the leading $3N$ force, which can be described by the well-known Fujita-Miyazawa diagram~\cite{Fujita:1957zz}, which involves intermediate excitation of the $\Delta$-isobars between three nucleons. 
While this contribution has to be considered in the $\Delta$-full approach, it can be shown that the net contribution of $3N$ forces vanishes in the delta-less chiral EFT~\cite{Weinberg:1992yk,vanKolck:1994yi} at this order.

At next order, N$^2$LO, the sub-leading $N\!N$ TPE diagrams contain vertices (large solid dots) proportional to the so-called $c_i$ coefficients. 
The values of these parameters can be obtained by pion-nucleon ($\pi N$)~\cite{Krebs:2007rh,Ditsche:2012fv,Hoferichter:2015dsa,Hoferichter:2015hva,
Hoferichter:2015tha,Siemens:2016hdi,Siemens:2016jwj,Yao:2016vbz} or $N\!N$ scattering data~\cite{Machleidt:2011zz}. 
In the delta-less chiral EFT, these coefficients mimic the effect of the $\Delta$-isobar (or some other meson resonances) through a mechanism known as resonance saturation. Hence, they are enhanced in magnitude and found to be "unnaturally" large. 
The explicit inclusion of the $\Delta$ isobar in the delta-full theory reduces the strength of the $c_i$'s and promotes the corresponding contributions to a lower order (see gray arrows in Fig.~\ref{fig:PowerCounting}).
As a consequence, the convergence of the expansion in the delta-full theory improves considerably at these orders. 
In the delta-full approach, additional sub-leading TPE contributions appear that have also been worked out at this order~\cite{Krebs:2007rh}. 

In addition to the $N\!N$ sector, additional $3N$ diagrams appear at N$^2$LO in both approaches. They involve a $3N$ TPE, a OPE-contact interaction, and a true $3N$ contact diagram. 
The $3N$ TPE potential also involves the $c_i$ parameters already present in the TPE $N\!N$ force. As in the case of the $N\!N$ force, these contributions absorb the presence of the $\Delta$-isobar in the delta-less approach, while some of their strength is promoted to lower order in the form of the already discussed Fujita-Miyazawa diagram in the delta-full approach. 
The OPE-contact and $3N$ contact diagrams include two purely three-body LECs that have to be adjusted to $A\geq 3$ data. Finally, the are no additional diagrams due to $\Delta$ contributions to the $3N$ force at N2LO~\cite{Epelbaum:2007sq}.

At higher orders, the number of contributions to the $N\!N$ force dramatically increases. In Fig.~\ref{fig:PowerCounting} only a few representative diagrams are displayed.
For instance, at N$^3$LO more TPE contributions occur -- in both delta-less and delta-full chiral EFT -- involving leading two-loop and relativistic corrections. In addition, leading three-pion ($3\pi$) exchange contributions arise at this order but they are found to be negligible. 
The main feature at N$^3$LO is the presence of additional contact interactions represented by the four-nucleon-leg with a solid diamond. Since these interactions are $\sim p^4, p'^4$, they have a relevant impact up to the $D$ waves. 
Their full operator structure will be discussed in the next section. 
Additional complicated $3N$ diagrams appear at N$^3$LO, as well as the first contributions to four-nucleon forces ($4N$). We will not discuss these diagrams here and refer the reader to Refs.~\cite{Bernard:2007sp,Bernard:2011zr,Epelbaum:2006eu,Epelbaum:2007us}.
For additional contributions at N$^4$LO and N$^5$LO, we refer the interested reader to Refs.~\cite{Entem:2015xwa,Epelbaum:2014sza,Reinert:2017usi,Entem:2017gor}.

An important aspect of nuclear interactions (and currents) in the basic model is that they suffer from ultraviolet (UV) divergences which need to be removed by a proper regularization and renormalization procedure. 
There are two sources of UV divergences that require regularization: first, UV divergences appear in loop corrections, and second when solving the Schr\"{o}dinger or Lippmann-Schwinger equations or when calculating matrix elements involving nuclear currents. 
Loop divergences can be treated via dimensional regularization (DR) or spectral-function regularization (SFR), where the latter is implemented through the inclusion of a finite cutoff in the spectral functions. 
To cure divergences when solving the Schr\"{o}dinger or Lippmann-Schwinger equations, the nuclear potential is multiplied by regulator functions that remove large-momentum contributions above a chosen cutoff scale.  
The regularization of the potential (and current) operators is followed by a renormalization procedure, i.e., dependencies on the regularization scheme and cutoff are reabsorbed, order by order, by the LECs entering the potential (and currents).

Nucleon-nucleon scattering has been extensively studied in chiral EFT in the past two decades following the pioneering work by Weinberg~\cite{Weinberg:1990rz,Weinberg:1991um,Weinberg:1992yk} and Ordonez et al.~\cite{Ordonez:1995rz}. 
In particular, $N\!N$ potentials at N$^3$LO in the chiral expansion are available since the early 2000's~\cite{Entem:2003ft,Epelbaum:2004fk} and have served as a basis for numerous {\it ab initio} calculations of nuclear structure and reactions.
Recently, accurate and precise chiral EFT potentials up to fifth order in the chiral expansion, i.e. N$^4$LO, have been developed~\cite{Entem:2015xwa,Epelbaum:2014sza,Reinert:2017usi,Entem:2017gor}, and provide an extremely accurate description of $N\!N$ data bases up to laboratory energies of 300 MeV with a $\chi^2$ per datum close to one. 
These databases have been provided by the Nijmegen group~\cite{Stoks:1993tb,Stoks:1994wp}, the VPI/GWU group~\cite{Arndt:2007qn}, and more recently the Granada group~\cite{Perez:2013jpa,Perez:2013oba,Perez:2014yla}. 
In the standard optimization procedure, the $N\!N$ potentials are first constrained through fits to neutron-proton ($np$) and proton-proton ($pp$) phase shifts, and then refined by minimizing the total $\chi^2$ obtained from a direct comparison with the $N\!N$ scattering data. 
However, new optimization schemes are being explored in Refs.~\cite{Carlsson:2015vda,Ekstrom:2015rta}.
For instance, the optimization strategy of the N2LO$_{\rm sat}$ interaction of Ref.~\cite{Ekstrom:2015rta} is based on a simultaneous fit of the two- and three-nucleon forces to low-energy $N\!N$ data, the deuteron binding energy, and binding energies and charge radii of hydrogen, helium, carbon, and oxygen isotopes using consistent $N\!N$ and $3N$ interactions at N$^2$LO. 
However, despite the good description of properties of $^{16}$O and $^{40}$Ca, the $N\!N$ component of this interaction shows deficiencies in reproducing the $pp$ and $np$ scattering data even at very low energy.

Three-nucleon forces and their impact on nuclear structure and reactions has become an important frontier in nuclear physics, see Refs.~\cite{KalantarNayestanaki:2011wz,Hammer:2012id} for review articles.
As shown in Fig.~\ref{fig:PowerCounting}, chiral contributions to the $3N$ interaction have been derived up to N4LO in the chiral expansion~\cite{Bernard:2007sp,Bernard:2011zr,Krebs:2012yv,Krebs:2013kha,Girlanda:2011fh}. 
However, few- and many-nucleon calculations are, with very few exceptions, still limited to chiral $3N$ forces at N$^2$LO. At this order, as we have mentioned above, $3N$ forces are characterized by the presence of two unknown LECs that have to be determined.
%they are given by a two-pion exchange contribution, whose strength is characteried by LECs fixed in the $N\!N$ or $\pi N$ sectors, and an OPE–contact interaction and a pure $3N$ contact contribution, with two unknown LECs to be determined. 
The two LECs -- namely $c_D$ in the OPE-contact and $c_E$ in the $3N$ contact interaction -- have been constrained either by fitting exclusively strong-interaction observables~\cite{Tews:2015ufa,Lynn:2015jua,Lynn:2017fxg,Piarulli:2017dwd} or by relying on a combination of strong- and weak- interaction observables~\cite{Gazit:2008ma,Marcucci:2011jm,Baroni:2018fdn}. 
This last approach is made possible by the relation between $c_D$ in the OPE-contact interaction and the LEC in the $N\!N$ contact axial current~\cite{Gazit:2008ma,Marcucci:2011jm}, established in chiral EFT~\cite{Gardestig:2006hj}. This connection allows one to use nuclear properties governed by either strong or weak interactions to constrain simultaneously the $3N$ interaction and $N\!N$ axial current.

%Within chiral EFT many studies have been carried out in the strong-interaction sector dealing with the construction and optimization of $N\!N$ and $3N$ interactions, and accompanying isospin symmetry-breaking corrections ~\cite{Ordonez:1995rz,Kaiser:1997mw,Kaiser:1998wa,Epelbaum:1998ka,Epelbaum:1999dj,Kaiser:1999ff,Kaiser:1999jg,Kaiser:2001dm,Kaiser:2001pc,Kaiser:2001at,Entem:2003ft,Machleidt:2011zz,Krebs:2007rh,vanKolck:1994yi,Epelbaum:2002vt,Navratil:2007zn,Bernard:2011zr,Girlanda:2011fh,Krebs:2012yv,Epelbaum:2014efa,Epelbaum:2014sza,Entem:2014msa,Entem:2015xwa,Ekstrom:2013kea,Ekstrom:2015rta,Ekstrom:2017koy,Reinert:2017usi,Binder:2018pgl,Krebs:2018jkc,Friar:1999zr,Friar:2004ca,Friar:2004rg}. Recently, two-nucleon interactions have been worked out up to sixth order of chiral perturbation theory and three-nucleon forces up to fifth order. 

As chiral EFT is a low-momentum expansion of nuclear interactions, many of the chiral interactions available in the literature are formulated in momentum space and have the feature of being strongly non-local in coordinate space. 
This makes them not well-suited for certain numerical algorithms, for example QMC methods. 
In this context, an interaction is local if it depends solely on the momentum transfer $\textbf{q}={\bf p} -{\bf p}^\prime$, which Fourier transforms to dependencies on $\textbf{r}$. 
However, interactions in momentum-space can also depend on the momentum scale ${\bf k}=({\bf p}^\prime +{\bf p})/2$, which Fourier transform to derivatives in coordinate space. 
These ${\bf k}$ dependencies, and thus non-localities, come about because of (i) the specific functional choice made to regularize the momentum space potentials in terms of the two momentum scales $\textbf{p}$ and $\textbf{p}'$, and (ii) contact interactions that explicitly depend on $\textbf{k}$.

QMC methods, for example variational (VMC) and Green's Function Monte Carlo (GFMC)~\cite{Carlson:2014vla,Lynn:2019rdt} techniques, provide reliable solutions of the many-body Schr\"odinger equation -- presently for up to $A\,$=$\,$12 nucleons -- with full account of the complexity of the many-body, spin- and isospin-dependent correlations induced by nuclear interactions. 
The sampling of configuration space in VMC and GFMC simulations gives access to many important properties of light nuclei such as spectra, form factors, transitions, low-energy scattering, and response functions. 
Auxiliary Field Diffusion Monte Carlo (AFDMC)~\cite{Carlson:2014vla,Lynn:2019rdt} uses Monte Carlo techniques to additionally sample the spin-isospin degrees of freedom, enabling studies of, for example, nuclei up to $A\,$=$\,$16~\cite{Lonardoni:2018nob,Lonardoni:2018sqo} and neutron matter~\cite{Tews:2015ufa,Lynn:2015jua,Tews:2018kmu,Tews:2018chv,Piarulli:2019pfq} that is so critical to determining the structure of neutron stars.
QMC simulations have surely proved to be very valuable in attacking many nuclear-structure problems over the last three decades but require local chiral interactions as input. 
Therefore, there is a need to develop local chiral interactions for the use in QMC methods in order combine these accurate many-body methods with systematic nuclear interactions and to test to what extent the chiral EFT framework impacts our knowledge of few- and many-body systems.

%\begin{equation}
%    V(\textbf{q})=-\frac{g_A^2}{(2f_{\pi})^2}\frac{\sigma_1\cdot \textbf{q}\, \sigma_2\cdot\textbf{q}}{q^2+m_{\pi}^2}\tau_1 \cdot \tau_2\,
%\end{equation}
%and in coordinate-space by 
%\begin{equation}
%    V(\textbf{r})= \,.
%\end{equation}

\section{Local Hamiltonians}
\label{sec:local}
\subsection{Local two-nucleon interactions}

A major thrust of our work is based on the theoretical derivation, optimization, and implementation of chiral interactions suitable for QMC methods. 
In recent years, local configuration-space chiral $N\!N$ interactions have been derived by two groups~\cite{Gezerlis:2013ipa,Gezerlis:2014zia,Piarulli:2014bda,Piarulli:2016vel}. 
In this section, we will introduce these two families of interactions, that are either derived in the delta-less~\cite{Gezerlis:2013ipa,Gezerlis:2014zia} or delta-full~\cite{Piarulli:2014bda,Piarulli:2016vel} approach. 
We begin by introducing general features of both approaches and then describe the specifics. 
We will be stating general considerations in momentum-space, where $\textbf{q}$ dependencies indicate local parts of interactions and $\textbf{k}$ dependencies indicate non-localities, and then switch to coordinate-space where interactions are local if they only depend on the relative distance $\textbf{r}=\textbf{r}_i-\textbf{r}_j$. 
Fourier transformations connect interactions in momentum- and coordinate-space, with $\textbf{q}$ and $\textbf{r}$ being associated variables, while $\textbf{k}$ leads to appearances of gradient terms.

As discussed before, nuclear interactions can generally be separated into different interaction channels depending on their operator structure. 
Obviously, chiral interactions can also be separated into long-range physics, mediated by pion-exchange interactions, and short-range physics, which is described by a set of operators consistent with all symmetries and accompanied by LECs adjusted to reproduce experimental data:
\begin{align}
    V({\bf q},{\bf k})=V_{\rm{cont}}({\bf q},{\bf k})+V_{\pi}({\bf q},{\bf k})\,.
\end{align}
Each of these components can then be expanded in chiral order $\nu$ as discussed before:
\begin{equation}
V_{i} = \sum_{\nu} V_i^{(\nu)}=V_i^{(0)} + V_i^{(2)} + V_i^{(3)} + V_i^{(4)} + \dots \,.
\label{eq:ch_pot}
\end{equation}

At LO, $\nu=0$, both delta-less and delta-full chiral EFT have the same operator structure. At this order, only the leading contact interactions as well as the one-pion exchange (OPE) interaction contribute, see Fig.~\ref{fig:PowerCounting}. Generally, pion-exchange interaction can be written as
\begin{align}
V_{\pi}&=V_{C,\pi}+  {\bm \tau}_i\cdot  {\bm \tau}_j W_C + \left(V_S+  {\bm \tau}_i\cdot  {\bm \tau}_j W_S \right)  {\bm \sigma}_i\cdot  {\bm \sigma}_j + \left(V_T+  {\bm \tau}_i\cdot  {\bm \tau}_j W_T \right)  {\bm \sigma}_i\cdot {\bf q} \,  {\bm \sigma}_j\cdot {\bf q}\\ \nonumber
&\quad + \left(V_{LS}+  {\bm \tau}_i\cdot  {\bm \tau}_j W_{LS} \right) i( {\bm \sigma}_i+{\bm \sigma}_j)\cdot {\bf q}\times {\bf k} + \left(V_{\sigma L}+  {\bm \tau}_i\cdot  {\bm \tau}_j W_{\sigma L} \right)  {\bm \sigma}_i\cdot {\bf q}\times {\bf k} \,\,\,  {\bm \sigma}_j\cdot {\bf q}\times {\bf k}\,,
\label{eq:PEops}
\end{align} 
with central, spin, tensor, spin-orbit and quadratic spin-orbit components, 
respectively. In the local chiral interactions discussed in this review, the spin-orbit and quadratic spin-orbit terms are not included as they are of higher order.
The one-pion exchange interaction is given in momentum space as
\begin{align}
V_{\text{OPE}}^{(0)}(\textbf{q})=-\frac{g_A^2}{4f_{\pi}^2}\frac{ {\bm \sigma}_i\cdot {\bf q}  {\bm \sigma}_j\cdot \bf{q}}{q^2+m_{\pi}^2} {\bm \tau}_i\cdot  {\bm \tau}_j\,,
\end{align}
where $g_A$, $f_\pi=92.4$ MeV, and $m_\pi$ denote the axial-vector coupling constant of the nucleon, the pion decay constant, and the pion mass, respectively. As a consequence, the OPE contributes to the $W_T$ channel. 

Including isospin-symmetry breaking effects induced by the mass difference between charged and neutral pions, the OPE interaction can be rewritten as
\begin{equation}
V_{\text{OPE}}^{(0)}(\textbf{q})=\left[ v^{\pi,\rm LO}_{\sigma \tau}(q) \, {\bm \sigma}_i \cdot {\bm \sigma}_j +
v^{\pi,\rm LO}_{t \tau}(q) \, S_{ij}({\bf q}) \right] \,  {\bm \tau}_i \cdot {\bm \tau}_j+\left[ v^{\pi,\rm LO}_{\sigma T}(q) \, {\bm \sigma}_i \cdot {\bm \sigma}_j +
v^{\pi,\rm LO}_{t T}(q) \, S_{ij}({\bf q}) \right]\,T_{ij}\,,
\label{eq:vpi}
\end{equation}
with the tensor operator $S_{ij}({\bf q})$ in momentum space, $S_{ij}({\bf q})= 3\, {\bm \sigma}_i \cdot {\bf q} \,\, {\bm \sigma}_j \cdot {\bf q}-q^2\, {\bm \sigma}_i \cdot {\bm \sigma}_j$, and the isotensor operator $T_{ij}=3\,\tau_{iz}\tau_{jz}-{\bm \tau}_i\cdot {\bm \tau}_j$. 
Hence, when including isospin-symmetry breaking, the OPE adds to the $W_S$ and $W_T$ parts of Eq.~\eqref{eq:PEops}.
The  functions, $v^{\pi,\rm LO}_{\sigma \tau}(q)$, $v^{\pi,\rm LO}_{t \tau}(q)$, $v^{\pi,\rm LO}_{\sigma T}(q)$, and $v^{\pi,\rm LO}_{t T}(q)$ are defined as 
\begin{eqnarray}
\label{eq:OPEfuncs}
v^{\pi,\rm LO}_{\sigma\tau}(q)&=&\frac{Y_0(q)+2\, Y_+(q)}{3}\,, \quad
v^{\pi,\rm LO}_{t\tau}(q)=\frac{T_0(q)+2\, T_+(q)}{3} \ ,\\
v^{\pi,\rm LO}_{\sigma T}(q)&=&\frac{Y_0(q)- Y_+(q)}{3}\,, \quad 
v^{\pi,\rm LO}_{tT}(q)=\frac{T_0(q)-T_+(q)}{3} \ ,\nonumber
\end{eqnarray}
with $Y_\alpha(q)$ and $T_\alpha(q)$ given by
\begin{equation}
Y_{\alpha}(q)= -\frac{g_A^2}{3\, (2f_{\pi})^2} \frac{q^2}{q^2+m_{\pi_\alpha}^2}\ ,\quad
T_{\alpha}(q)=-\frac{g_A^2}{3\, (2f_{\pi})^2} \frac{1}{q^2+m_{\pi_\alpha}^2} \ .
\end{equation}
Here, $m_{\pi_\alpha}$ denotes the neutral ($m_{\pi_0}$) and charged ($m_{\pi_{\pm}}$) pion masses. 
When Fourier-transformed, the coordinate-space OPE is given by
\begin{equation}
\label{eq:vpi_r}
 v^{\pi,\rm LO}({\bf r})=\left[ v^{\pi,\rm LO}_{\sigma \tau}(r) \, {\bm \sigma}_i \cdot {\bm \sigma}_j +
v^{\pi,\rm LO}_{t \tau}(r) \, S_{ij}({\bf r})\right] \,  {\bm \tau}_i \cdot {\bm \tau}_j+\left[ v^{\pi,\rm LO}_{\sigma T}(r) \, {\bm \sigma}_i \cdot {\bm \sigma}_j +
v^{\pi,\rm LO}_{t T}(r) \, S_{ij}\right]\,T_{ij}\,
\end{equation}
where the individual functions can be obtained from Eq.~\eqref{eq:OPEfuncs} with $q\to r$ and with the functions $Y_\alpha(r)$ and $T_\alpha(r)$ given by
\begin{equation}
Y_\alpha(r)= \frac{g_A^2}{12\, \pi}\,  \frac{m^3_{\pi_\alpha}}{(2\,f_{\pi})^2} \,
\frac{e^{-x_\alpha}}{x_\alpha}\ ,\quad
T_\alpha(r)=Y_\alpha(r)\left( 1+\frac{3}{x_\alpha}+\frac{3}{x^2_\alpha}\right)\ .
\end{equation}
Here, $x_\alpha=m_{\pi_\alpha} r$.
Note that Eq.~(\ref{eq:vpi_r}) only holds in the case $r>0$. In addition, upon Fourier transformation a $\delta$-function appears, which has been dropped from Eq.~(\ref{eq:vpi_r}), because it can be reabsorbed in the short-range contact terms at LO, which we will discuss next. 

The LO contact interactions are momentum-independent and can be described by the most general operator set allowed by all symmetries:
\begin{align}
    V_{\rm cont}^{\rm LO}({\bf q},{\bf k})=V_{\rm cont}^{\rm LO}=\alpha_1 \mathbbm{1}+ \alpha_2\,\bm\sigma_i\cdot\bm\sigma_j+\alpha_3\,\bm\tau_i\cdot\bm\tau_j+ \alpha_4\,\bm\sigma_i\cdot\bm\sigma_j\,\bm\tau_i\cdot\bm\tau_j\,.
    \label{eq:LOpot}
\end{align}
As these terms describe the interactions of nucleons, i.e., fermions, these interactions are used between antisymmetrized wave functions. One can define the antisymmetrized interaction $V_{\text{as}}=1/2 \left(V-\mathcal{A}[V] \right)$
by applying the antisymmetrizer, given by 
\begin{align}
\mathcal{A}[V({\bf q},{\bf k})]&=\frac14 (1+{\bm \sigma}_i \cdot {\bm
\sigma}_j)(1+{\bm \tau}_i \cdot {\bm \tau}_j) \times V\left({\bf q} \rightarrow -2{\bf k}, {\bf k} \rightarrow -\frac12 {\bf q}\right)\,.
\label{eq:antisymmetrizer}
\end{align}
One then finds
\begin{align}
V^{(0)}_{\text{cont,as}} &= \frac12 \left(1- \frac14 (1+{\bm \sigma}_i \cdot {\bm \sigma}_j)(1+{\bm \tau}_i \cdot {\bm \tau}_j)\right) V_ {\text{cont}}^{(0)} \nonumber \\
&= \left(\frac{3}{8}\alpha_1 -\frac{3}{8}\alpha_2 -\frac{3}{8}\alpha_3 -\frac{9}{8}\alpha_4 \right) + \left(-\frac{1}{8}\alpha_1 +\frac{5}{8}\alpha_2 -\frac{3}{8}\alpha_3 +\frac{3}{8}\alpha_4 \right) { \bm \sigma}_i \cdot { \bm \sigma}_j \nonumber \\
& \quad + \left(-\frac{1}{8}\alpha_1 -\frac{3}{8}\alpha_2 +\frac{5}{8}\alpha_3 +\frac{3}{8}\alpha_4 \right) {\bm \tau}_i \cdot {\bm \tau}_j + \left(-\frac{1}{8}\alpha_1 +\frac{1}{8}\alpha_2 +\frac{1}{8}\alpha_3 +\frac{3}{8}\alpha_4 \right) { \bm\sigma}_i \cdot { \bm\sigma}_j \, {\bm \tau}_i \cdot {\bm \tau}_j \nonumber \\
&=\tilde C_S + \tilde C_T \,{\bm \sigma}_i \cdot {\bm \sigma}_j + \left(-\frac{2}{3}\tilde C_S-\tilde C_T \right) {\bm \tau}_i \cdot {\bm \tau}_j + \left(-\frac13 \tilde C_S\right) {\bm \sigma}_i \cdot { \bm\sigma}_j \, {\bm \tau}_i \cdot {\bm \tau}_j \,.
\label{eq:LO_antisymm}
\end{align}
It follows immediately that only two out of these four couplings are linearly independent, describing the two possible $S$-wave scattering channels. The two commonly chosen LO contact operators are
\begin{equation}
V^{(0)}_{\text{cont}} = C_S + C_T {\bm \sigma}_i \cdot {\bm \sigma}_j\,,
\label{eq:antisym_LO}
\end{equation}
but in principle any different two of the four contact interactions can be chosen and lead to the same physical description for fermionic systems. This is analogous to Fierz ambiguities and in the following we will call this freedom to choose operators Fierz rearrangement freedom. 

Additionally, there are isospin breaking corrections to the LO contact interactions that have to be taken into account. These are due to different masses of u and d quarks, and account for differences in neutron-neutron ($nn$), $np$, and $pp$ $S$-wave scattering lengths:
\begin{align}
\label{cont_IB}
V_{\rm cont, \; CIB} ({\bf r}) &= C_{\rm CIB} \frac{1 + 4 {\bm \tau}_i^3 {\bm \tau}_j^3 }{2} \frac{1-  {\bm \sigma}_i \cdot {\bm \sigma}_k}{4}\,, \\
V_{\rm cont, \; CSB} ({\bf r}) &= C_{\rm CSB} ({\bm \tau}_i^3 + {\bm \tau}_j^3 ) \frac{1- {\bm  \sigma}_i \cdot {\bm \sigma}_k}{4}
\,.\label{cont_IB2}
 \end{align}

At higher orders, the description of the potential changes depending on the choice of delta-less or delta-full approach. In the following, we will describe both approaches as pursued by individual research groups.

\subsubsection{Without Delta isobars}

At NLO in chiral EFT, additional momentum-dependent contact interactions as well as TPE interactions appear. For the TPE, we give the expressions within the spectral-function representation (SFR) as detailed in Ref.~\cite{Epelbaum:2003gr}, with spectral functions $\rho_i$ and $\eta_i$:
\begin{align}
V_{C,\pi} (r) &= \frac{1}{2 \pi^2 r} \int_{2 M_\pi}^{\tilde \Lambda} d \mu \, \mu \, 
e^{-\mu r} \, \rho_{C} (\mu) \,, \label{four11} \\
V_{S} (r) &= -\frac{1}{6 \pi^2 r} \int_{2 M_\pi}^{\tilde \Lambda} d \mu \, \mu \, 
e^{-\mu r} \, \Bigl( \mu^2 \rho_T (\mu) - 3 \rho_S (\mu ) \Bigr)\, , \label{four12} \\
V_{T} (r) &= -\frac{1}{6 \pi^2 r^3} \int_{2 M_\pi}^{\tilde \Lambda}  d \mu \, \mu \, 
e^{-\mu r} \, \rho_T (\mu) \, ( 3 + 3 \mu r + \mu^2 r^2 )\,. 
\label{four13}
\end{align}
Here, $\tilde \Lambda$ is the SFR cutoff. Similar expressions are valid for $W_{C}$, $W_{S}$, and $W_{T}$ in terms of $\eta_{C}$, $\eta_S$, and $\eta_T$. The TPE spectral functions at NLO are given by~\cite{Kaiser:1997mw} 
\begin{align}
\rho_{T}^{(2)} (\mu) &= \frac{1}{\mu^2}\, \rho_{S}^{(2)} (\mu) 
= \frac{3 g_A^4}{128 \pi f_\pi^4} \,
\frac{\sqrt{\mu^2 - 4 m_\pi^2}}{\mu} \label{spectr_nlo1}
\,, \\
\eta_{C}^{(2)} (\mu) &= 
\frac{ 1}{768 \pi f_\pi^4}\,
\frac{\sqrt{\mu^2 - 4 m_\pi^2}}{\mu} \biggl(4m_\pi^2 (5g_A^4 - 4g_A^2 - 1) - \mu^2(23g_A^4 - 10g_A^2 -1) 
+ \frac{48 g_A^4 m_\pi^4}{4 m_\pi^2 - \mu^2} \biggr) \label{spectr_nlo2} \,.
\end{align}
For the NLO contact interactions, the most general set of operators is given by
\begin{align}
    V_ {\rm cont}^{\rm NLO}({\bf q},{\bf k}) = & \gamma_1 \, q^2 + \gamma_2 \, q^2\, {\bm \sigma}_i \cdot {\bm \sigma}_j + \gamma_3 \, q^2\, {\bm \tau}_i \cdot {\bm \tau}_j  + \gamma_4 \, q^2 {\bm \sigma}_i \cdot {\bm \sigma}_j {\bm \tau}_i \cdot {\bm \tau}_j + \gamma_5 \, k^2 + \gamma_6 \, k^2\, {\bm \sigma}_i \cdot {\bm \sigma}_j \nonumber \\
    + & \gamma_7 \, k^2\, {\bm \tau}_i \cdot {\bm \tau}_j + \gamma_8 \, k^2 {\bm \sigma}_i \cdot {\bm \sigma}_j {\bm \tau}_i \cdot {\bm \tau}_j + \gamma_9 \, ({\bm \sigma}_i + {\bm \sigma}_j)({\bf q}\times {\bf k}) + \gamma_{10} \, ({\bm \sigma}_i + {\bm \sigma}_j)({\bf q}\times {\bf k}) {\bm \tau}_i \cdot {\bm \tau}_j \nonumber \\
    + & \gamma_{11} ({\bm \sigma}_i \cdot {\bf q}) ({\bm \sigma}_j \cdot {\bf q}) + \gamma_{12} ({\bm \sigma}_i \cdot {\bf q}) ({\bm \sigma}_j \cdot {\bf q}) {\bm \tau}_i \cdot {\bm \tau}_j  + \gamma_{13} ({\bm \sigma}_i \cdot {\bf k}) ({\bm \sigma}_j \cdot {\bf k}) \nonumber \\
    + & \gamma_{14} ({\bm \sigma}_i \cdot {\bf k}) ({\bm \sigma}_j \cdot {\bf k}) {\bm \tau}_i \cdot {\bm \tau}_j \,.
\end{align}
Using the same arguments as for the LO contact interactions, only 7 out of these 14 operators are linearly independent. To construct local interactions, one typically chooses the 6 local operators (proportional to $\gamma_1$-$\gamma_4$, $\gamma_{11}$, and $\gamma_{12}$) as well as the spin-orbit operator (proportional to $\gamma_9$):
\begin{align} 
V^{(2)}_{\rm cont} &= C_1 \, q^2 + C_2 \, q^2 \, 
{\bm \tau}_i \cdot {\bm \tau}_j  + \bigl(C_3 \, q^2 + C_4 \, q^2 \, {\bm \tau}_i \cdot {\bm \tau}_j \bigr)
\, {\bm \sigma}_i \cdot {\bm \sigma}_j + i \, \frac{C_5}{2} \, ({\bm \sigma}_i + {\bm \sigma}_j) \cdot
({\bf q} \times {\bf k}) \nonumber \\
&\quad + C_6 \, ({\bm \sigma}_i \cdot {\bf q})({\bm \sigma}_j \cdot {\bf q}) 
 + C_7 \, ({\bm \sigma}_i \cdot {\bf q})({\bm \sigma}_j \cdot {\bf q}) 
\, {\bm \tau}_i \cdot {\bm \tau}_j \,.
\label{eq:NLOshort}
\end{align}
In coordinate space, this translates to 
\begin{align}
V^{(2)}_{\rm cont}({\bf r}) &= -(C_1+C_2 \,{\bm \tau}_i \cdot {\bm
\tau}_j)\Delta \delta ({\bf r}) -(C_3+C_4\, {\bm \tau}_i \cdot {\bm \tau}_j) \, {\bm \sigma}_i \cdot
{\bm \sigma}_j \Delta \delta ({\bf r}) + \frac{C_5}{2}\frac{\partial_r \delta ({\bf r})}{r} {\bf L}\cdot {\bf S} \nonumber \\
&\quad + (C_6+ C_7 \,{\bm \tau}_i \cdot {\bm \tau}_j) \times \left[({\bm \sigma}_i \cdot \hat{\bf r})({\bm \sigma}_j
\cdot \hat{\bf r}) \left(\frac{\partial_r \delta ({\bf
r})}{r}- \partial_r^2 \delta ({\bf r}) \right)- {\bm \sigma}_i \cdot {\bm \sigma}_j \frac{\partial_r \delta ({\bf r})}{r}  \right]\,.
\label{eq:NLO_FT}
\end{align}
At N$^2$LO, the subleading TPE interactions appear. The spectral functions for these at N$^2$LO read 
\begin{align}
\rho_{C}^{(3)} (\mu) &= -\frac{3 g_A^2}{64 \mu f_\pi^4} \,(2m_\pi^2 - \mu^2) \, \Big(2m_\pi^2(2c_1 -c_3) + c_3 \mu^2 \Big) 
\,, \label{spectr_nnlo1}\\ 
\eta_{T}^{(3)} (\mu) &= \frac{1}{\mu^2}\, \eta_{S}^{(3)} (\mu) =
- \frac{g_A^2}{128 \mu f_\pi^4} \, c_4 (4m_\pi^2 - \mu^2)
\,,\label{spectr_nnlo2}
\end{align}
where the $c_i$ denote the previously mentioned LECs of the subleading pion-nucleon vertices. 
For the N$^2$LO TPE, one can solve Eqs.~(\ref{four11})--(\ref{four13}):
\begin{align}
\label{coord_nnlo2}
W_S^{(3)} (r) &=  \frac{g_A^2}{48 \pi^2 f_\pi^4} \, \frac{e^{- 2 x}}{r^6}
c_4 \, (1 + x) (3 + 3 x + 2 x^2)\nonumber  \\
&\quad - {}  \frac{g_A^2}{384 \pi^2 f_\pi^4} \, \frac{e^{- y}}{r^6}
c_4 \, \Big( 24 + 24 y + 12 y^2 + 4 y^3 + y^4  - 4 x^2 ( 2 + 2 y + y^2) \Big)\, ,
\end{align}
\begin{align}
W_T^{(3)} (r) &= - \frac{g_A^2}{48 \pi^2 f_\pi^4} \, \frac{e^{- 2 x}}{r^6}
c_4 \, (1 + x) (3 + 3 x + x^2)  \nonumber\\
& \quad + {}  \frac{g_A^2}{768 \pi^2 f_\pi^4} \, \frac{e^{- y}}{r^6}
c_4 \, \Big( 48 + 48 y + 24 y^2 + 7 y^3 + y^4 - 4 x^2 ( 8 + 5 y + y^2) \Big)\,,\label{coord_nnlo3}
\end{align}
and
\begin{align}
\label{coord_nnlo1}
V_{C,\pi}^{(3)} (r) &=  \frac{3 g_A^2}{32 \pi^2 f_\pi^4} \, \frac{e^{- 2 x}}{r^6}
\bigg[ 2 c_1 \, x^2 (1 + x)^2 + c_3 (6 + 12x + 10 x^2 + 4 x^3 + x^4) \bigg] \nonumber \\
& \quad- {} \frac{3 g_A^2}{128 \pi^2 f_\pi^4} \, \frac{e^{- y}}{r^6} \bigg[
4 c_1 x^2 \Big(2 + y ( 2 + y) - 2 x^2   \Big) + c_3 \Big(  24 +  y (24 + 12 y  + 4 y^2 + y^3   )\nonumber \\
&\quad - 4 x^2 (2 + 2 y  + y^2 ) + 4  x^4 \Big) \bigg]\,,
\end{align}
where $x \equiv m_\pi r$ and $y \equiv \tilde \Lambda r$. 

The relativistic $1/m_N$ corrections, with $m_N$ being the nucleon mass, have been omitted here since, in the counting employed here, they would appear at N$^3$LO, provided the nucleon mass is counted according to $Q/m_N \sim Q^2/\Lambda_b^2$ as suggested in Ref.~\cite{Weinberg:1991um}.

The delta-less chiral EFT approach has been used to construct local interactions up to N$^2$LO. At next higher order, N$^3$LO, contact interactions cannot be written down in a purely local fashion, as only 8 out of 30 possible operators are local. A possible way forward is the definition of ''maximally local'' N$^3$LO potentials, which has been pursued in the delta-full approach and will be discussed in the next section.

\begin{figure}[t]
\begin{center}
\includegraphics[width=0.48\textwidth]{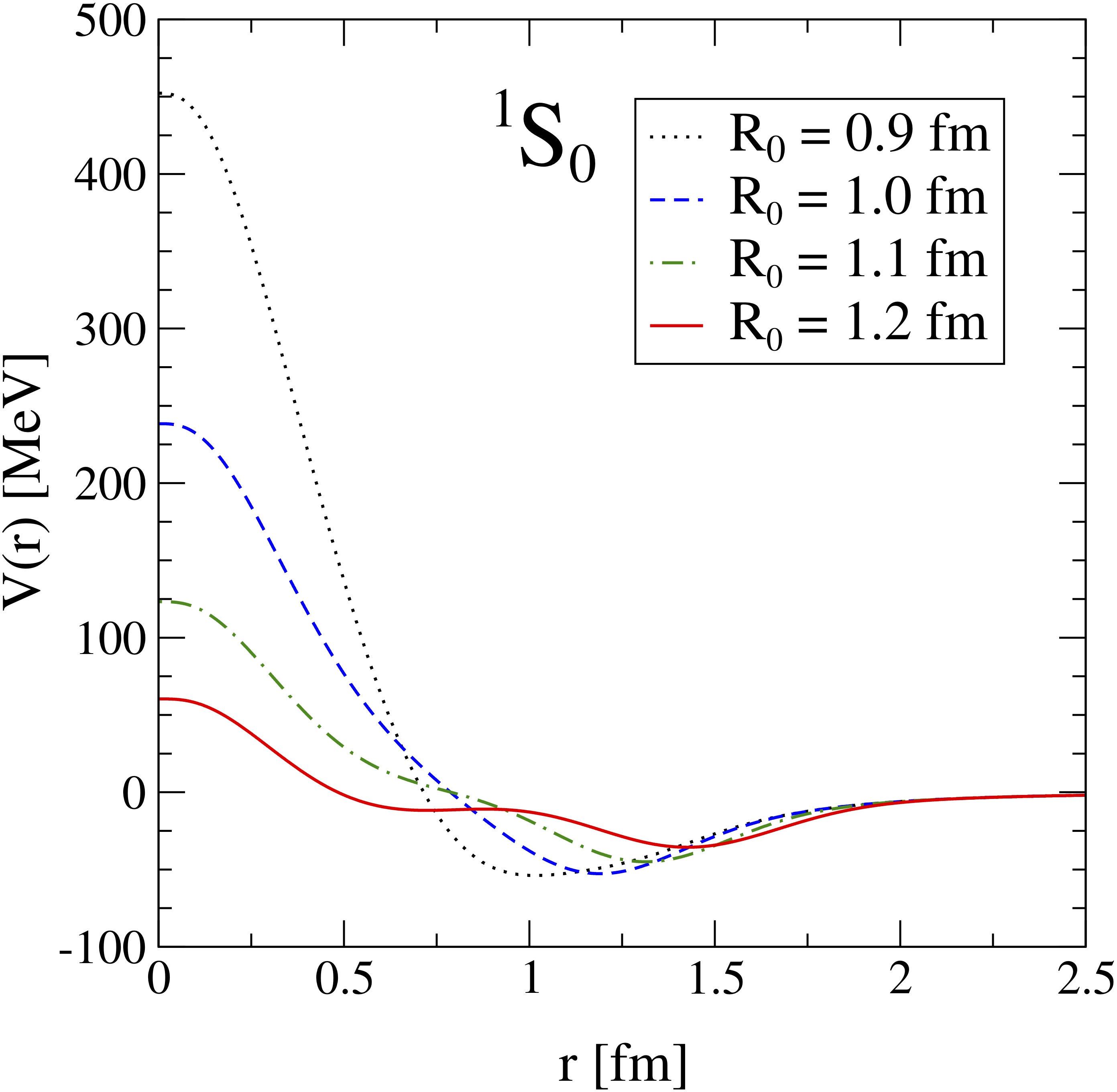}
\end{center}
\caption{Local delta-less chiral potential in the $^1$S$_0$ partial wave at N$^2$LO. The smaller the coordinate-space cutoff $R_0$, the smaller is the short-range repulsive core. Figure taken from Ref.~\cite{Gezerlis:2014zia} under the Creative Commons CCBY license.}\label{fig:pot}
\end{figure}

Finally, it is necessary to specify a regulator scheme. For the delta-less local interactions of Refs.~\cite{Gezerlis:2013ipa,Gezerlis:2014zia}, the following long- and short-range regulators are used:
\begin{align}
    f_{\rm{long}}(r)&=\left(1- e^{ -\left(\frac{r}{R_0} \right)^{n_1}} \right)^{n_2} \,,\quad 
    f_{\rm{short}}(r)=\frac{n}{4 \pi\,R_0^3\,\Gamma\left(\frac{3}{n}\right)}\,e^{-\left(\frac{r}{R_0} \right)^{n}} \,. \label{eq:fshort}
\end{align} 

The long-range regulator multiplies each function $Y(r)$, while the short-range regulator replaces all $\delta$ functions. The regulator functions depend on the cutoff scale $R_0$, that determines how long- and short-range physics are separated. For a smaller cutoff $R_0$ (i.e., for a larger momentum-space cutoff), the interactions is probed at shorter distances, and typically shows stronger short-range repulsion. We show the delta-less local chiral interactions in the $^1$S$_0$ channel in Fig.~\ref{fig:pot} for different values of the cutoff. Introducing a local regulator function leads to the appearance of regulator artifacts that brake Fierz-rearrangement freedom. We will address this topic in detail in Section~\ref{sec:Fierz}.

\subsubsection{With Delta isobars}

In the delta-full local chiral interactions, coordinate-space expressions for the TPE terms at NLO and N$^2$LO are obtained by using the spectral function representation~\cite{Kaiser:1997mw,Epelbaum:2003gr} but with dimensional regularization (DR)~\cite{Kaiser:1998wa}. 
This implies taking the cutoff $\tilde \Lambda$ in Eqs.~(\ref{four11})--(\ref{four13}) to infinity ($\tilde \Lambda \rightarrow \infty$).
Consequently, the terms depending on the variable $y$ in Eqs.~(\ref{coord_nnlo1})--(\ref{coord_nnlo2}) vanish. 
For the relevant radial functions involved in the one- and two-delta diagrams up to N$^2$LO, we refer the interested reader to Appendix A. 
The singularities at the origin of the OPE and TPE components are regularized by cutoff functions of the form
\begin{equation}
\label{eq:ctff}
 f_{\rm long}^{\Delta}(r)=1-\frac{1}{(r/R_{\rm L})^6 \,  e^{(r-R_{\rm L})/a_{\rm L}} +1} \ , 
\end{equation}
where three values for the radius $R_{\rm L}$ are considered: $R_{\rm L}=(0.8,1.0,1.2)$ fm with the diffuseness $a_{\rm L}$ fixed at $a_{\rm L}=R_{\rm L}/2$ in each case. 

Another difference between the delta-less and delta-full coordinate-space interactions lies in the operator structure of their short-range components.
In the delta-full potentials, selected contact terms at N$^3$LO are also retained in addition to the LO and NLO contributions given in Eqs.~(\ref{eq:antisym_LO}) and~(\ref{eq:NLO_FT}).
The contact potential at order N3LO, $V^{\rm N3LO}_{\rm cont}(q,k)$, which involves four gradients acting on the nucleon fields, is expressed in terms of 15 independent operators~\cite{Machleidt:2011zz} after considering the Fierz rearrangement freedom. Its standard parametrization, adopted in momentum-space potentials, is given by 
\begin{align}
 \label{eq:vct4}
V^{\rm N3LO}_{\rm cont}(q,k)&=\tilde{D}_1\,q^4+\tilde{D}_2\,k^4
+\tilde{D}_3\,q^2\,k^2+\tilde{D}_4\,({\bf k} \times {\bf q})^2+\Big[\tilde{D}_5\,q^4+
\tilde{D}_6\,k^4+\tilde{D}_7\,q^2 k^2\nonumber\\
&+\tilde{D}_8\,({\bf k} \times {\bf q})^2\Big]\,{\bm \sigma}_i\cdot {\bm \sigma}_j+i\,(\tilde{D}_9\,q^2+\tilde{D}_{10}\,k^2)
\,{\bf S}\cdot \left({\bf k} \times {\bf q}\right)+(\tilde{D}_{11}\,q^2+\tilde{D}_{12}\,k^2)\,S_{ij}({\bf k})\nonumber\\
&+(\tilde{D}_{13}\,q^2+\tilde{D}_{14}\,k^2)\,S_{ij}({\bf k})+{\tilde D}_{15}\,[{\bm \sigma}_i\cdot ({\bf k}\times {\bf q})\,{\bm \sigma}_j\cdot ({\bf k}\times {\bf q})]\ .
\end{align}
However terms proportional to $k^2$ and $k^4$ in those expressions, upon Fourier transformation, would lead to gradient operators in coordinate-space (${\bf p}\longrightarrow-i{\bm \nabla}$ is the
relative momentum operator), making the $N\!N$ potential strongly non-local.

The number of non-localities can be reduced by reconsidering the Fierz rearrangement freedom. However some of these non-local terms still persist at N$^3$LO leading to the definition of "minimally nonlocal" contact interactions:
\begin{align}
V^{\rm N3LO}_{\rm cont}(q,k)&=D_1\,q^4+D_2\,q^4\,{\bm \tau}_i\cdot {\bm \tau}_j+D_3\,q^4\,{\bm \sigma}_i\cdot {\bm \sigma}_j+D_4\,q^4\,{\bm \sigma}_i\cdot {\bm \sigma}_j\,{\bm \tau}_i\cdot {\bm \tau}_j+D_5\,q^2\, S_{ij}({\bf q}) \nonumber\\
&+D_6\,q^2\, S_{ij}({\bf q})\,{\bm \tau}_i\cdot {\bm \tau}_j
+iD_7\,q^2\,{\bf S}\cdot \left({\bf k} \times {\bf q}\right)+i\,D_8\,q^2\,{\bf S}\cdot \left({\bf k}\, \times {\bf q}\right){\bm \tau}_i\cdot {\bm \tau}_j\nonumber\\
&+D_{9}\left[{\bf S}\cdot \left({\bf k} \times {\bf q}\right)\right]^2+D_{10}\left({\bf k} \times {\bf q}\right)^2+D_{11}\left({\bf k} \times {\bf q}\right)^2{\bm \sigma}_i\cdot {\bm \sigma}_j+D_{12}\,q^2k^2\nonumber\\
&+D_{13}\,q^2k^2{\bm \sigma}_i\cdot {\bm \sigma}_j+D_{14}\,k^2\,  S_{ij}({\bf q})+D_{15}\,k^2\, S_{ij}({\bf q})
\,{\bm \tau}_i\cdot {\bm \tau}_j
\label{eq:sci2}\ .
\end{align}
In coordinate space, this reads as 
\begin{align}
\label{eq:vr}
V^{(3)}_{\rm cont}(\bf r)&=\left[\sum_{l=1}^{11} v_{\rm S}^l (r) \, O^l_{ij}\right]+\{\,v_{\rm S}^p(r)
+v_{\rm S}^{p\sigma}(r)\,{\bm \sigma}_i\cdot {\bm \sigma}_j+v_{\rm S}^{pt}(r)\,S_{ij}(\textbf{r})+
v_{\rm S}^{pt\tau}(r)\,S_{ij}(\textbf{r})\,{\bm \tau}_i\cdot {\bm \tau}_j\,\, ,\,\,{\bf p}^2\,\}\,,
\end{align}
where 
\begin{equation}
O^{l=1,\dots,11}_{ij}=\lbrace {\bf 1}\, ,\, {\bm \tau}_i\cdot {\bm \tau}_j\, ,\,{\bm \sigma}_i\cdot {\bm \sigma}_j\, ,\, {\bm \sigma}_i\cdot {\bm \sigma}_j\,{\bm \tau}_i\cdot {\bm \tau}_j\, ,
S_{ij}\, ,\,S_{ij}\,{\bm \tau}_i\cdot {\bm \tau}_j\, ,\, {\bf L}\cdot{\bf S}\,,\,
 {\bf L}\cdot{\bf S}\,{\bm \tau}_i\cdot {\bm \tau}_j\, ,\, ({\bf L}\cdot{\bf S})^2\, ,\, {\bf L}^2\, ,\, 
 {\bf L}^2\, {\bm \sigma}_i\cdot {\bm \sigma}_j \rbrace\ ,
 \end{equation}
referred to as $c$,  $\tau$, $\sigma$, $\sigma \tau$, $t$, $t\tau$ for the first six operators, and $b$, $b\tau$, $bb$, $q$, $q\sigma$ for the remaining five operators.
 The four additional terms, denoted as $p$, $p\sigma$, $pt$, and $pt\tau$, in the
anti-commutator of Eq.~(\ref{eq:vr}) are ${\bf p}^2$-dependent. For the definition of the radial functions $v_{\rm S}^l (r)$ as well as those multiplying the ${\bf p}^2$-terms, we refer the reader to Appendix A.

A comment is now in order. The strict adherence to power counting would require the inclusion of additional one-loop as well as two-loop TPE and three-pion exchange contributions at N$^3$LO. For the time being, these contributions
	have been neglected, since part of their strength is promoted at lower orders due to the inclusion of the $\Delta$ resonance, and some of the remaining diagrams are also known to be small (see, for example, Ref.~\cite{Machleidt:2011zz}). Furthermore it is the $D_i$ LEC's at
	N$^3$LO that are critical for a good reproduction of phase shifts in lower partial waves, particularly $D$-waves, and a good fit to the $N\!N$ database. However, the consistency between the long- and short-range part at higher orders in the delta-full chiral EFT is work in progress.

The local versions of these "minimally nonlocal" $N\!N$ potentials have been defined by dropping the terms proportional to ${\bf p}^2$ in the anti-commutator when the optimization procedure for estimating the LECs is carried out~\cite{Piarulli:2016vel}. 
In Ref.~\cite{Piarulli:2016vel} we observed that the inclusion of the ${\bf p}^2$-dependent terms would have improved the fits to the database in the laboratory energy range up to 200 MeV only marginally. 
However, apart from the small improvement that the ${\bf p}^2$-dependent terms
would bring to the total $\chi^2$ in the fit to the $N\!N$ scattering data, the effect of these terms on nuclear observables has not been studied.

Lastly, the delta-full local interactions contain additional isospin breaking terms at NLO. They are parametrized by the following operators
\begin{equation}
O^{l=12,\dots,16}_{ij}= \{\tau_i^z+\tau_j^z,\,T_{ij},\,{\bm \sigma}_i\cdot {\bm \sigma}_j T_{ij},\,S_{ij}\,T_{ij},\,{\bf L}\cdot{\bf S}\,T_{ij} \}\ ,
 \end{equation}
referred to as  $\tau z$, $T$, $\sigma T$, $tT$, $bT$. The radial functions multiplying these operators are also reported in Appendix A.

The short-range part of these potentials involve
the local regulator given in Eq.~(\ref{eq:fshort}) with $n=2$,
\begin{equation}
f_{\rm short}^{\Delta}(r)=\frac{1}{\pi^{3/2}R_{\rm S}^3} e^{-(r/R_{\rm S})^2} ,
\end{equation}
where we consider, in combination with $R_{\rm L}=(0.8,1.0,1.2)$ fm,
$R_{\rm S}=(0.6,0.7,0.8)$ fm, corresponding to typical momentum-space
cutoffs $\Lambda_{\rm S}=2/R_{\rm S}$ ranging from about 660 MeV down to 500 MeV.
Hereafter, we will denote the potential with cutoffs $(R_{\rm L},R_{\rm S})=(1.2,0.8)$ fm
as model $a$, that with $(1.0,0.7)$ fm as model $b$, and that with $(0.8,0.6)$ fm as
model $c$.  

There are 26 LECs in the definition of the delta-full local interactions. 
Of these, 20 LECs describe the charge-independent part of the interaction: 2 at LO ($Q^0$), 7 at NLO ($Q^2$), and 11 at N$^3$LO ($Q^4$). 
The remaining 6 LECs describe its charge-dependent part: 2 at LO (one each from CIB and CSB), and 4 at NLO from CIB. 
The optimization procedure to fix these 26 LECs uses $pp$ and $np$ scattering data (including normalizations), as assembled in the Granada database~\cite{Perez:2013jpa}, the $N\!N$ scattering length, and the deuteron binding energy. 
The minimization of the $\chi^2$ objective function with respect to the LECs is carried out with the Practical Optimization Using no Derivatives (for Squares) routine, POUNDerS~\cite{PhysRevC.82.024313}.
For each of three different sets of cutoff radii $(R_{\rm S},R_{\rm L})$, two classes of local interactions have been developed, which only differ in the range of laboratory energy over which the fits were carried out, either 0--125 MeV in class I or 0--200 MeV in class II. 
The $\chi^2$/datum achieved by the fits in class I (II) was $\lesssim 1.1(\lesssim1.4)$ for a total of about 2700 (3700) data points. 
In the literature, we are referring to these $N\!N$ interactions generically as the Norfolk potentials (NV2s), and designate those in class I as NV2-Ia, NV2-Ib, and NV2-Ic, and those in class II as NV2-IIa, NV2-IIb, and NV2-IIc. 

The NV2 interactions were found to provide insufficient attraction in calculations 
of the ground-state energies of nuclei with $A\,$=$\, 3$--6~\cite{Piarulli:2016vel}. To remedy this and similar shortcomings, $3N$ interactions at N$^2$LO have to be included in both approaches. This will be described in the next section.

\subsection{Local three-nucleon interactions}

Three-nucleon forces are very important ingredients for the correct description of physical systems.
They naturally appear within chiral EFT and are consistent with the $N\!N$ sector. The exact description of the $3N$ interactions depends on the choice of delta-less vs. delta-full approach. In the following, we review $3N$ forces in both approaches.

\subsubsection{Without Delta isobars}

In the delta-less chiral EFT approach, the leading $3N$ contributions appear at N$^2$LO in the power counting. 
They an be separated into three topologies: (i) a long-range TPE interaction named $V_C$ depending on the pion-nucleon LECs $c_1$, $c_3$, and $c_4$, that already appear in the $N\!N$ sector, (ii) a one-pion-exchange--contact interaction $V_D$ dependent on a new LEC $c_D$, and (iii) a 3N contact interaction $V_E$ dependent in a new LEC $c_E$. The LECs $c_D$ and $c_E$ solely describe $3N$ physics and need to be adjusted to properties of $A \geq 3$ systems.
In momentum space, these interactions are defined as
\begin{align}
V_C &= \frac{1}{2}\left(\frac{g_A}{2 f_{\pi}}\right)^2 \sum_{\pi(ijk)} \frac{{\bm \sigma}_i\cdot {\bf{q}}_i \, {\bm \sigma}_k\cdot {\bf{q}}_k}{(q_i^2+m_{\pi}^2)(q_k^2+m_{\pi}^2)} \, F_{ijk}^{\alpha \beta} \, {\bm \tau}_i^{\alpha} \, {\bm \tau}_k^{\beta} \,, \label{eq:Vc} \\
V_D &= -\frac{g_A}{8f_{\pi}^2}\frac{c_D}{f_{\pi}^2 \Lambda_{\chi}} \sum_{\pi(ijk)} \frac{{\bm \sigma}_k\cdot {\bf{q}}_k}{q_k^2+m_{\pi}^2} \, {\bm \sigma}_i \cdot {\bf{q}}_k \, { \bm \tau}_i \cdot { \bm \tau}_k \,, \label{eq:Vd} \\
V_E &= \frac{c_E}{2 f_{\pi}^4 \Lambda_{\chi}}  \sum_{\pi(ijk)} { \bm \tau}_i \cdot { \bm \tau}_k \,, \label{eq:Ve}
\end{align}
where we sum over all permutations of the particles $i$, $j$, and $k$, where the first pion carries a momentum ${\bf q}_i$ from nucleon i to j, while the second pion carries ${\bf q}_k$ from j to k, and where $F_{ijk}^{\alpha \beta}$ is given by
\begin{align}
F_{ijk}^{\alpha \beta}&=\delta^{\alpha \beta}\left[-\frac{4c_1m_{\pi}^2}{f_{\pi}^2}+\frac{2c_3}{f_{\pi}^2} \, {\bf{q}}_i\cdot {\bf{q}}_k \right] +\sum_{\gamma} \frac{c_4}{f_{\pi}^2} \varepsilon^{\alpha \beta \gamma} \, {\bm \tau}_j^{\gamma} \, {\bm \sigma}_j \cdot ({\bf{q}}_i \times {\bf{q}}_k) \,.
\end{align}

As one can easily see, all of these interactions are local, as long as local regulator functions are applied. To obtain expressions in coordinate space, these interactions have to be Fourier transformed. For the part of $V_C$ proportional to $c_1$, we find
\begin{align}
V_{C,c_1}^{ijk} &= -\frac{c_1m_{\pi}^2 g_A^2}{2f_{\pi}^4}
\sum_{\pi(ijk)} { \bm \tau}_i \cdot { \bm \tau}_k \int \frac{d^3 q_i}{(2\pi)^3} \frac{{\bm \sigma}_i\cdot \textbf{q}_i}{q_i^2+m_{\pi}^2} \, e^{i \textbf{q}_i \cdot \textbf{r}_{ij}} \, \int\frac{d^3q_k}{(2\pi)^3} \frac{{\bm \sigma}_k\cdot \textbf{q}_k}{q_k^2+m_{\pi}^2} \, e^{i \textbf{q}_k \cdot \textbf{r}_{kj}} \,.
\end{align}
This results in 
\begin{align}
V_{C,c_1}^{ijk} &= \frac{c_1 m_{\pi}^6 g_A^2}{2 f_{\pi}^4 (4 \pi)^2}
\sum_{\pi(ijk)} { \bm \tau}_i \cdot { \bm \tau}_k \, {\bm \sigma}_i \cdot \hat{\textbf{r}}_{ij} \, {\bm \sigma}_k \cdot \hat{\textbf{r}}_{kj} \, U(r_{ij}) Y(r_{ij}) U(r_{kj}) Y(r_{kj}) \,,
\end{align}
where we have used
\begin{align}
\int \frac{d^3q_i}{(2\pi)^3} \frac{{\bm \sigma}_i\cdot \textbf{q}_i}{q_i^2+m_{\pi}^2} \, e^{i \textbf{q}_i \cdot \textbf{r}_{ij}} &= -i \, {\sigma}_i^{\alpha} \, \partial^{\alpha} \, \frac{e^{-m_{\pi} r_{ij}}}{4 \pi r_{ij}} = i \, \frac{m_\pi^2}{4 \pi} \, {\sigma}_i^{\alpha} \, \hat{r}_{ij}^\alpha \, U(r_{ij}) Y(r_{ij}) \,,
\end{align} 
and 
\begin{equation}
Y(r)=\frac{\exp(-m_{\pi}\cdot r)}{m_{\pi} r}\,, \quad U(r)=1+\frac{1}{m_{\pi} r}\,. \\
\end{equation}

For the other parts of $V_C$ we find
\begin{align}
\label{eq:compvcc1}
V_{C,c_3}&=\frac{g_A^2m_\pi^6c_3}{2304\pi^2f_\pi^4}
\sum_{\pi(ijk)}\{ { \bm \tau}_i \cdot { \bm \tau}_k, { \bm \tau}_k \cdot { \bm \tau}_j\}
\{\mathcal{X}_{ik}({\bf r}_{ik}),\mathcal{X}_{kj}({\bf r}_{kj})\}\,,\\
\label{eq:compvcc4}
V_{C,c_4}&=-\frac{g_A^2m_\pi^6c_4}{4608\pi^2f_\pi^4}
\sum_{\pi(ijk)}[ { \bm \tau}_i \cdot { \bm \tau}_k, { \bm \tau}_k \cdot { \bm \tau}_j]
[\mathcal{X}_{ik}({\bf r}_{ik}),\mathcal{X}_{kj}({\bf r}_{kj})]\,,
\end{align}
where
\begin{align}
\label{eq:mathcalX}
\mathcal{X}_{ij}({\bf r})= X_{ij}({\bf r})-\frac{4\pi}{m_\pi^3}  \delta({\bf r}){\bm \sigma}_i \cdot {\bm \sigma}_j\,,\quad 
X_{ij}({\bf r})&=\left(S_{ij}({\bf r})T(r) + {\bm \sigma}_i \cdot {\bm \sigma}_jY(r) \right)\,,
\end{align}
and 
\begin{eqnarray}
T(r)=\left( 1 + \frac{3}{{m_{\pi}}r} + \frac{3}{m_{\pi}^2\, r^2} \right) Y(r) \, .
\end{eqnarray}

For the one-pion-exchange--contact part $V_D$ we find
\begin{align}
\label{eq:compvd}
V_D^{ijk} &= \frac{g_A}{24f_{\pi}^2}\frac{c_D}{f_{\pi}^2 \Lambda_{\chi}}
\sum_{\pi(ijk)} { \bm \tau}_i\cdot { \bm \tau}_k \biggl[ \frac{m_{\pi}^3}{4 \pi} \, \delta(\textbf{r}_{ij}) X_{ik}(\textbf{r}_{kj}) - {\bm \sigma}_i \cdot {\bm \sigma}_k \, \delta(\textbf{r}_{ij}) \delta(\textbf{r}_{kj}) \biggr] \,,
\end{align}
and for the three-nucleon--contact interaction $V_E$ we find
\begin{align}
V_E&= \frac{c_E}{2 f_{\pi}^4 \Lambda_{\chi}} \sum_{\pi(ijk)} {\bm \tau}_i \cdot  {\bm \tau}_k\delta( \textbf{r}_{ij}) \delta( \textbf{r}_{kj}) \,.
\end{align}
To regularize these $3N$ topologies, we choose consistent regulators with the $N\!N$ sector, i.e., we replace $\delta$ functions by $f_{\rm short}(r)$ and multiply Yukawa functions with $f_{\rm long}(r)$. The cutoff scale for $3N$ interactions does not necessarily have to be the same as for the $N\!N$ sector, and we call it $R_{\text{3N}}$ in the following. 

\begin{figure}[t]
\begin{center}
\includegraphics[width=0.45\textwidth]{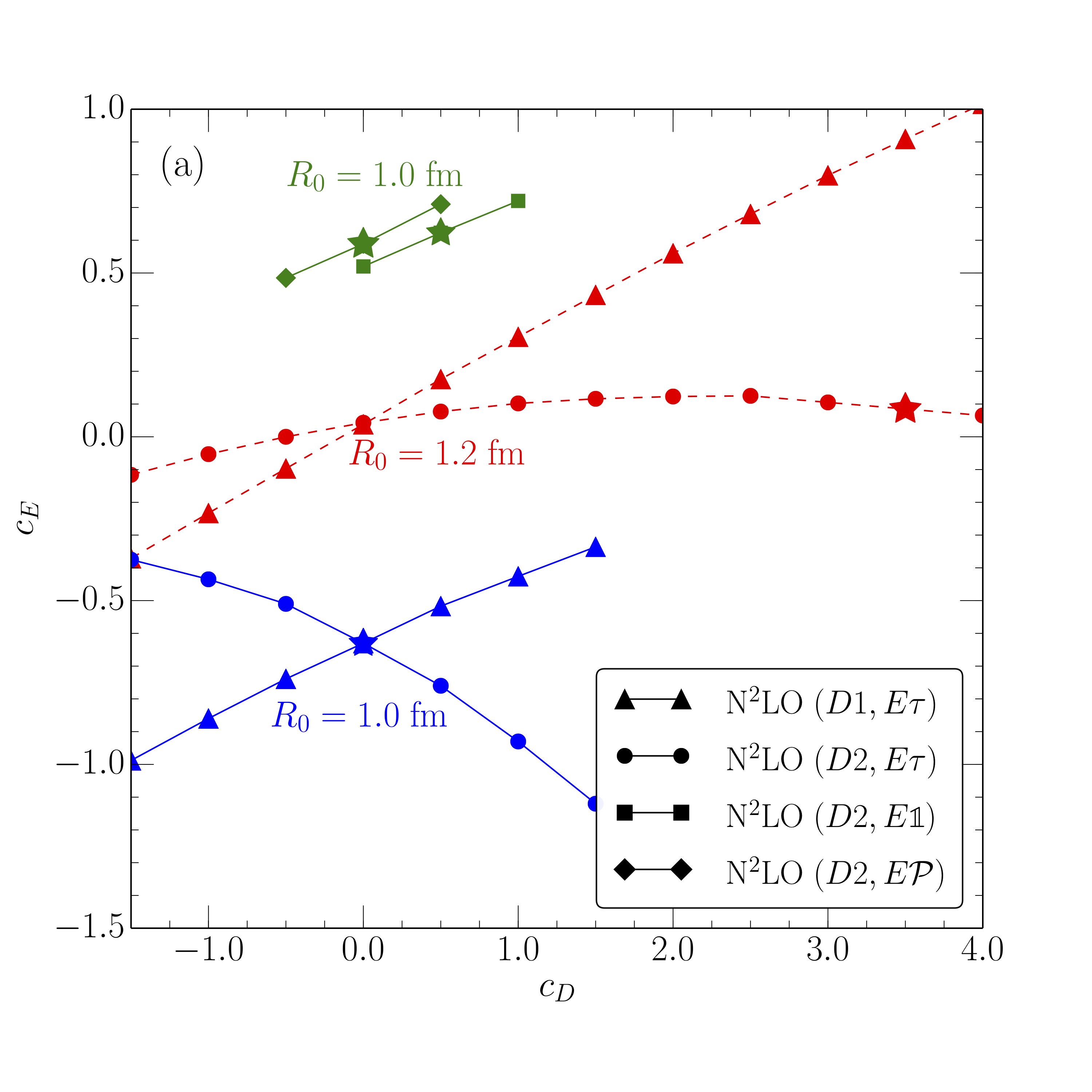}
\includegraphics[width=0.45\textwidth]{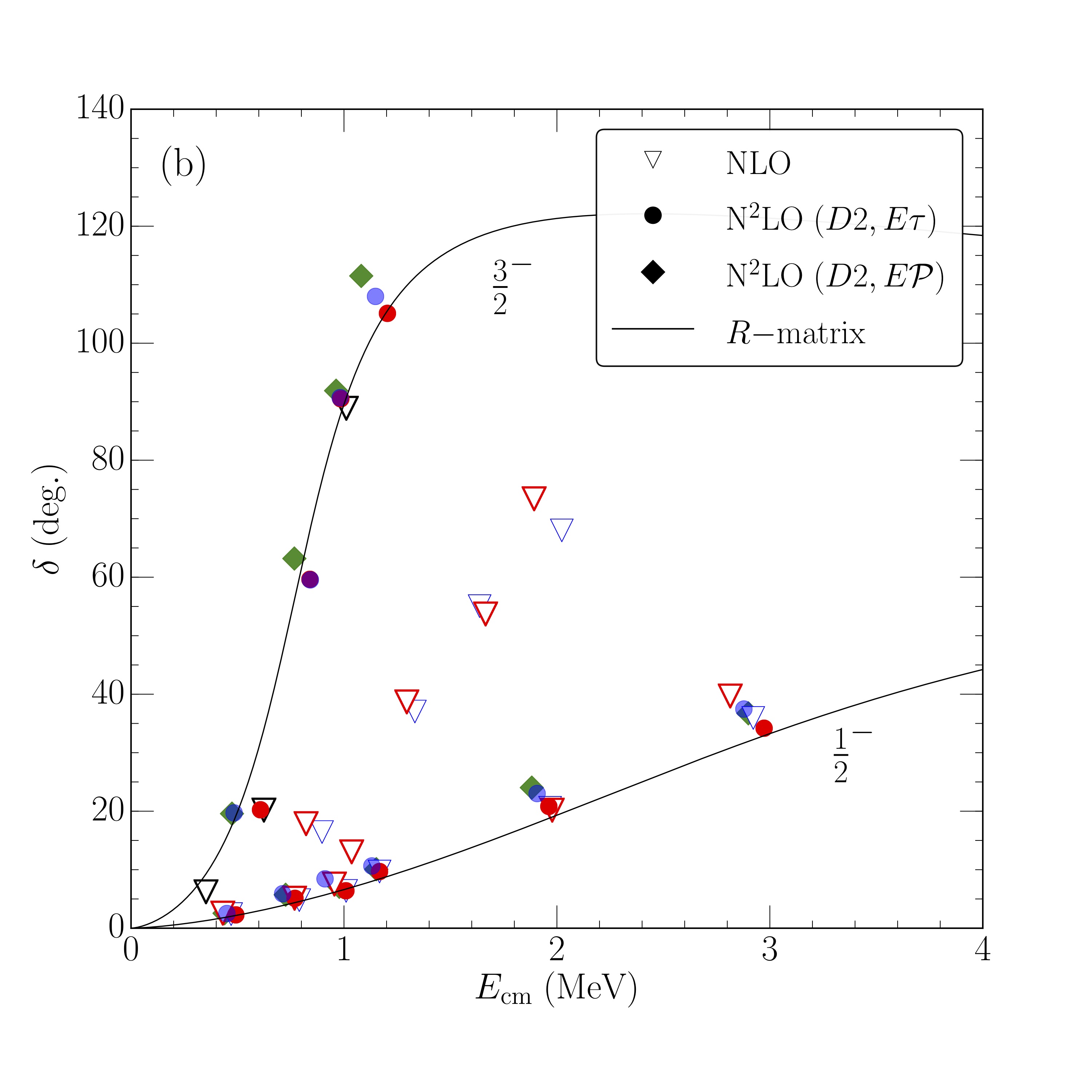}
\end{center}
\caption{Left: Parameter curves for the LECs $c_D$ and $c_E$ for the binding energy of $^4$He for different cutoffs and $3N$ parametrizations discussed in Sec.~\ref{sec:reg_artifacts}. Right: Reproduction of n-$\alpha$ $P$ wave phase shifts at NLO and at N$^2$LO for the parameter combinations marked by a star in the left panel. Figures taken from Ref.~\cite{Lynn:2015jua} under the Creative Commons CCBY license.}\label{fig:3Nfits_Joel}
\end{figure}

To adjust the appearing $3N$ couplings to experimental data, one should select few-body observables that are uncorrelated. 
In the delta-less approach, these observables have been chosen to be the $^4$He binding energy and n-$\alpha$ scattering $P$ wave phase shifts; see Fig.~\ref{fig:3Nfits_Joel}, where we show parameter curves for the $3N$ LECs for different 3N cutoffs $R_{\text{3N}}$, chosen similar to $R_0$, and for different parametrizations that we will discuss in the next section. 
Stars in the parameter curves mark fits that also describe neutron-alpha scattering, shown in the right panel. For more details, see Ref.~\cite{Lynn:2015jua}.

\subsubsection{With Delta isobars}

In the delta-full chiral EFT approach, the structure of the $3N$ force at N$^2$LO is similar to the $3N$ force in the delta-less approach. We still have the three topologies $V_{C}$, $V_{D}$ and $V_{E}$ at N$^2$LO but, in addition, the well-known Fujita-Miyazawa interaction~\cite{Fujita:1957zz} ($V_{\Delta}$), which in the delta-less approach is absorbed by $V_C$, appears already at NLO in the power counting. In momentum space, it reads as
\begin{eqnarray}
\label{eq:e1}
V_{\Delta}^{ijk}
&=&-\frac{g_A^2\, h_A^2}{16\, f_\pi^4} \, \frac{1}{m_{\Delta N}\,  \, (q_i^2+m_{\pi}^2)( q_k^2+m_{\pi}^2)}\,\Big[
{\bm \sigma}_k \cdot {\bf q}_k \,\, {\bf S}^\dagger_j\cdot {\bf q}_k \,\,{\bf S}_j\cdot {\bf q}_i \, {\bm \sigma}_i \cdot {\bf q}_i\,\,
{\bm \tau}_k \cdot {\bf T}^\dagger_j \,\,{\bf T}_j\cdot  {\bm \tau}_i \\
&&-{\bm \sigma}_i \cdot {\bf q}_i \,\, {\bf S}^\dagger_j\cdot {\bf q}_i \,\,{\bf S}_j\cdot {\bf q}_k \, {\bm \sigma}_k \cdot {\bf q}_k\,\,
{\bm \tau}_i \cdot {\bf T}^\dagger_j \,\,{\bf T}_j\cdot  {\bm \tau}_k\Big] \ ,
\end{eqnarray}
where $S$, $S^\dagger$ and $T$, $T^\dagger$ are the transition spin and isospin operators: The operator $S$ ($T$) converts a spin (isospin) 1/2 into a spin (isospin) 3/2 particle. 
%Here, the pion “1" carries a momentum ${\bf q}_1$ from nucleon i to j, while “2" carries ${\bf q}_2$ from j to k. IT: moved that explanation to the previous section and will replace here with qij and qjk

The configuration-space expression follows from
\begin{eqnarray}
\label{eq:e6}
V_{\Delta}^{ijk}
&=& -\frac{g_A^2\, h_A^2 }{16\cdot 144\, \pi^2}\frac{m_\pi^6}{m_{\Delta N} f_\pi^4} 
\left[ X_{jk}^{II\, \dagger} X_{ji}^{II}  \,\,{\bf T}^\dagger_j \cdot {\bm \tau}_k\,\,{\bf T}_j\cdot  {\bm \tau}_i 
+X_{ji}^{II\, \dagger} X_{jk}^{II}  \,\,{\bf T}^\dagger_j \cdot {\bm \tau}_i\,\,{\bf T}_j\cdot  {\bm \tau}_k \right] \ ,
\end{eqnarray}
where the following definitions have been introduced:
\begin{eqnarray}
X^{II}_{ij} &=& T(r_{ij})\, S^{II}_{ij} + Y(r_{ij})\,  
{\bf S}_i \cdot {\bm \sigma}_j \,, \quad S^{II}_{ij} = 3\, {\bf S}_i \cdot \hat{\bf r}_{ij} \, {\bm \sigma}_j \cdot \hat{\bf r}_{ij}
-{\bf S}_i \cdot {\bm \sigma}_j \,
\end{eqnarray}
and the dimensionless functions $Y(r)$ and $T(r)$ defined before.
%given by
%\begin{eqnarray}
%Y(r)=\frac{e^{-{m_{\pi}}r}}{{m_{\pi}}r}\, , \qquad T(r)=\left( 1 + \frac{3}{{m_{\pi}}r} %+ \frac{3}{m_{\pi}^2\, r^2} \right) Y(r) \ .
%\end{eqnarray}
The term $\left[\, \cdots\, \right]$ in Eq.~(\ref{eq:e6}) can be written as
\begin{eqnarray}
\label{eq:3.23nr}
\left[ \cdots\right] =&&\frac{1}{2} \Big[
\left( X^{II\, \dagger}_{jk}\, X^{II}_{ji} +{\rm h.c.} \right)
\left( {\bf T}^{\dagger}_j \cdot {\bm \tau}_k\, {\bf T}_j \cdot {\bm \tau}_i
+{\rm h.c.}\right) \nonumber \\
&&+\left(X^{II \dagger}_{jk}\, X^{II}_{ji} -{\rm h.c.}\right)
\left({\bf T}^{\dagger}_j \cdot {\bm \tau}_k\, {\bf T}_j \cdot {\bm \tau}_i -{\rm h.c.}\right) \Big] ,
\end{eqnarray}
and the transition-spin and transition-isospin operators can be eliminated
using the identities 
\begin{eqnarray}
\label{eq:3.21nn}
X^{II \dagger}_{jk}\, X^{II}_{ji} + {\rm h.c.} &=&
\frac{2}{3} \Big\{ X_{jk}\, , \, X_{ji} \Big\} \ ,   \\
\label{eq:3.21n}
X^{II \dagger}_{jk}\, X^{II}_{ji} - {\rm h.c.} &=&
-\frac{1}{3} \Big[ X_{jk}\, , \, X_{ji} \Big] \ ,  \\ 
\label{eq:3.22n}
{\bf T}^{\dagger}_j \cdot {\bm \tau}_k\, {\bf T}_j \cdot {\bm \tau}_i + {\rm h.c.}
&=& \frac{2}{3} \Big\{ {\bm \tau}_j \cdot {\bm \tau}_k\, ,\, 
{\bm \tau}_j \cdot {\bm \tau}_i \Big\}\ , \\
\label{eq:3.21nx}
{\bf T}^{\dagger}_j \cdot {\bm \tau}_k \, {\bf T}_j \cdot {\bm \tau}_i - {\rm h.c.}
&=& -\frac{1}{3} \Big[{\bm \tau}_j \cdot {\bm \tau}_k\, ,\,
{\bm \tau}_j \cdot {\bm \tau}_i \Big] \ ,
\label{eq:3.22nx}
\end{eqnarray}
to obtain
\begin{equation}
\label{eq:3.25n}
V_{\Delta}^{ijk} =  -  \frac{g_A^2 h_A^2} 
{ 72\cdot 144\, \pi^2} \,\frac{m_\pi^6} {m_{\Delta N}f_\pi^4} \bigg[
\Big \{ {X}_{ij}\, ,\, {X}_{jk} \Big\} 
\Big\{ {\bm \tau}_i \cdot {\bm \tau}_j\, ,\, {\bm \tau}_j \cdot {\bm \tau}_k \Big\}+\frac{1}{4}\,
\Big[ {X}_{ij}\, ,\, {X}_{jk} \Big]
\Big[ {\bm \tau}_i \cdot {\bm \tau}_j \, ,\,  {\bm \tau}_j \cdot {\bm \tau}_k \Big]  \bigg]\,,
\end{equation}
where the function $X_{ij}$ was defined in the previous section. In the definitions above, the $\delta({\bf r})$-function terms have been dropped.
%Note that here the notation for $X_{ij}$ reads as
%\begin{equation}
%X_{ij} = T_{\pi}(r_{ij})\, S_{ij} + Y_{\pi}(r_{ij})\,
%{\bm \sigma}_i \cdot {\bm \sigma}_j \ .
%\end{equation}

\begin{figure}[t]
	\center
	\includegraphics[height=2.3in]{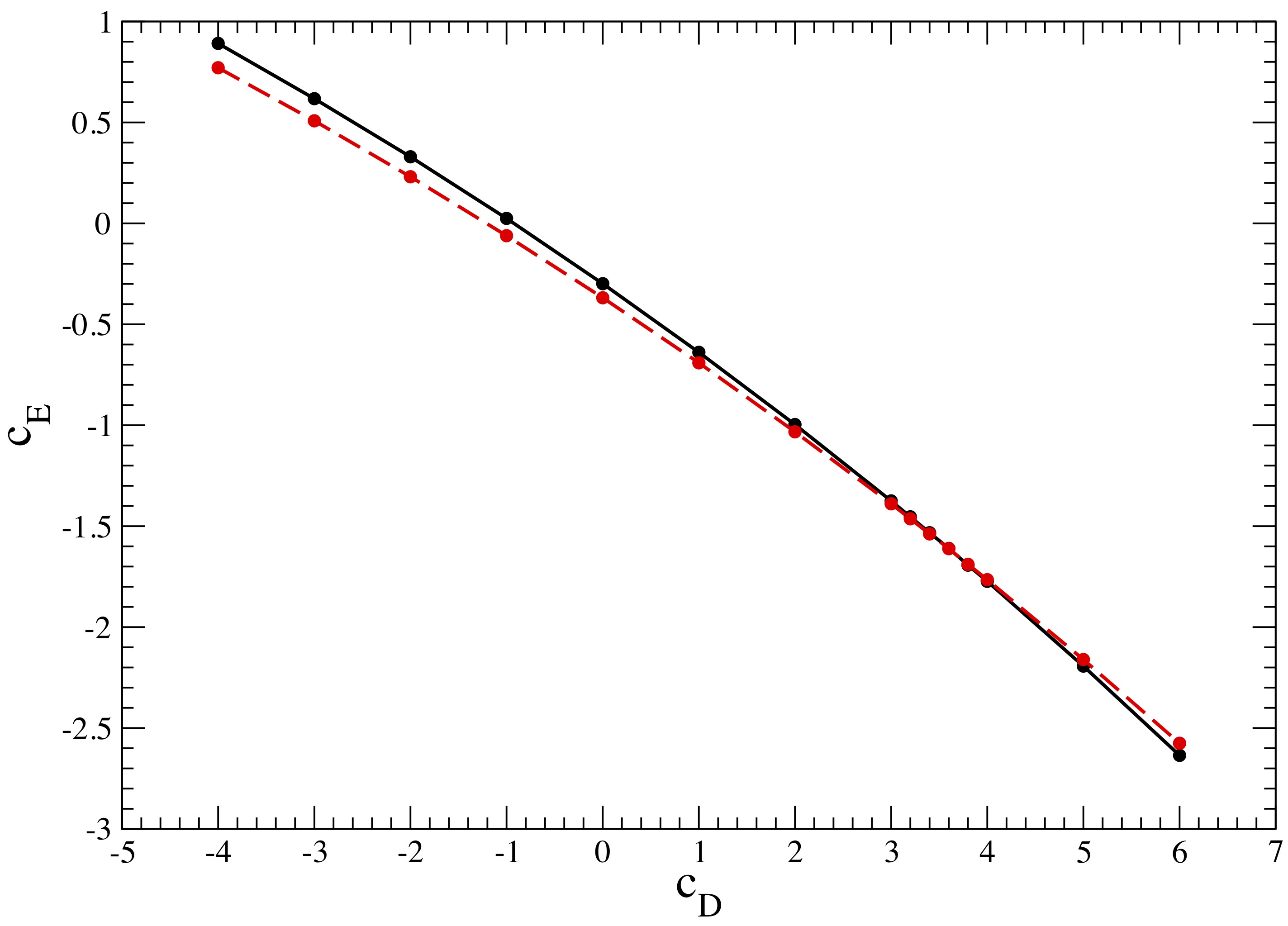} \vspace*{-2.5in}
	\includegraphics[height=2.3in]{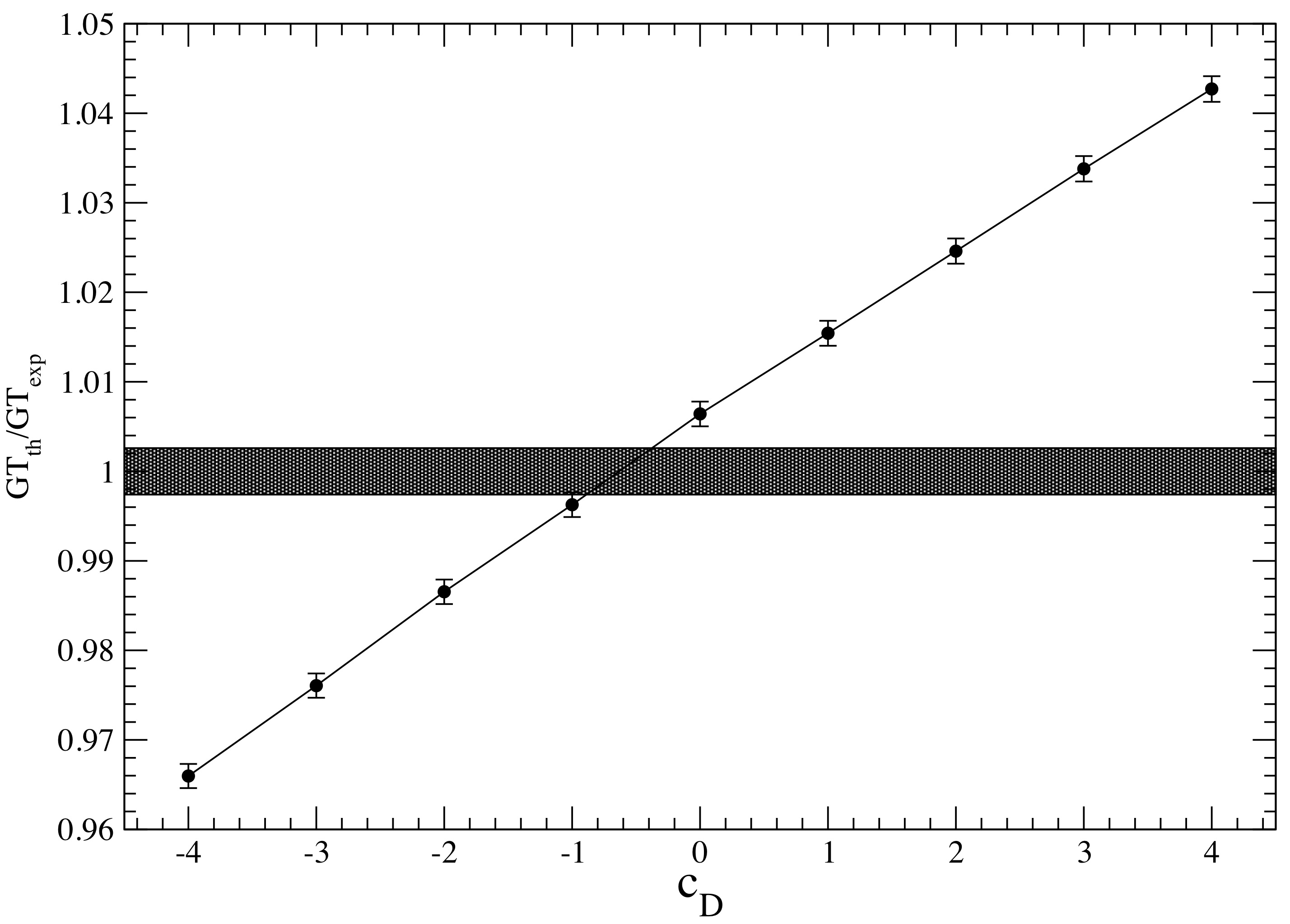} \vspace*{2.5in}
	\caption{Left panel: The $c_D$-$c_E$ trajectories obtained by fitting the
		experimental trinucleon binding energies (solid line) and $nd$ doublet scattering length (dashed line) (the intercept of 
		these two lines gives the $c_D$ and $c_E$ values that reproduce these two observables simultaneously). Figure taken from Ref.~\cite{Piarulli:2017dwd} under the Creative Commons CCBY license.
		Right panel: The calculated ratio GT$_{\rm th}$/GT$_{\rm exp}$ as function of $c_D$ (solid line; each point on 
		this line reproduces the trinucleon binding energies). Figure taken from Refs.~\cite{Baroni:2018fdn} under the Creative Commons CCBY license.} %The NV2+3-Ia chiral interactions are used here for illustration.
	\label{fig:fitting}
\end{figure}
In analogy to the $3N$ delta-less chiral EFT, we regularize the $3N$ contributions in the delta-full chiral EFT by replacing the $\delta$ functions with $f_{\rm short}^{\Delta}(r)$ and multiplying the Yukawa functions with $f_{\rm long}^{\Delta}(r)$. Note that the implementation of $V_C$ and $V_D$ in the delta-full chiral EFT does not retain the terms proportional to ${\bm \sigma}_i \cdot {\bm \sigma}_j$ in the definition of $\mathcal{X}_{ik}$, in Eq.~(\ref{eq:mathcalX}), and in Eq.~(\ref{eq:compvd}). They can be reabsorbed in the redefinition of the short-range contact terms.

In the delta-full chiral EFT, two different sets for the values of $c_D$ and $c_E$ were obtained, leading to two different parametrization of the $3N$ interaction~\cite{Piarulli:2017dwd,Baroni:2018fdn}. In the first, these LECs were determined by simultaneously reproducing the experimental trinucleon ground-state energies and neutron-deuteron ($nd$) doublet scattering length, as shown in the left panel of Fig.~\ref{fig:fitting}.
In the second set, these $c_D$ and $c_E$ were constrained by fitting, in addition to the trinucleon energies, the empirical value of the Gamow-Teller matrix element in tritium $\beta$ decay~\cite{Baroni:2018fdn}, see right panel of
Fig.~\ref{fig:fitting}. Because of the much reduced
correlation between binding energies and the GT matrix
element, the second fit procedure leads to a more robust determination of $c_D$ and $c_E$ then attained in the first one. Note that these observables have been calculated with hyperspherical-harmonics (HH) expansion methods~\cite{Kievsky:2008es} as described in Refs.~\cite{Piarulli:2017dwd,Baroni:2018fdn}.

%The resulting Hamiltonian models were designated as NV2+3-Ia/b and NV2+3-IIa/b (or Ia/b and IIa/b for short) in the first case, and as NV2+3-Ia$^*$/b$^*$ and NV2+3-IIa$^*$/b$^*$ (or Ia$^*$/b$^*$ and IIa$^*$/b$^*$) in the second.

\section{Finite cutoff and regulator artifacts}\label{sec:reg_artifacts}

The derivations of local interactions in the last sections did not include any of the local regulator functions that necessarily have to be applied to the interactions to make them suitable for the use in nuclear many-body methods. 
Generally, when introducing a regulator function, terms beyond the order at which one is working are affected. Hence, the use of such regulator functions with finite values for the cutoff leads to the appearance of regulator artifacts, that might influence calculations of many-body observables. In this section, we will address the different regulator artifacts that can appear in calculations with local interactions.   

\begin{figure}[t]
\centering
\vspace*{-0.5cm}
\includegraphics[trim=0 0 30cm 0, clip=,width=0.45\textwidth]{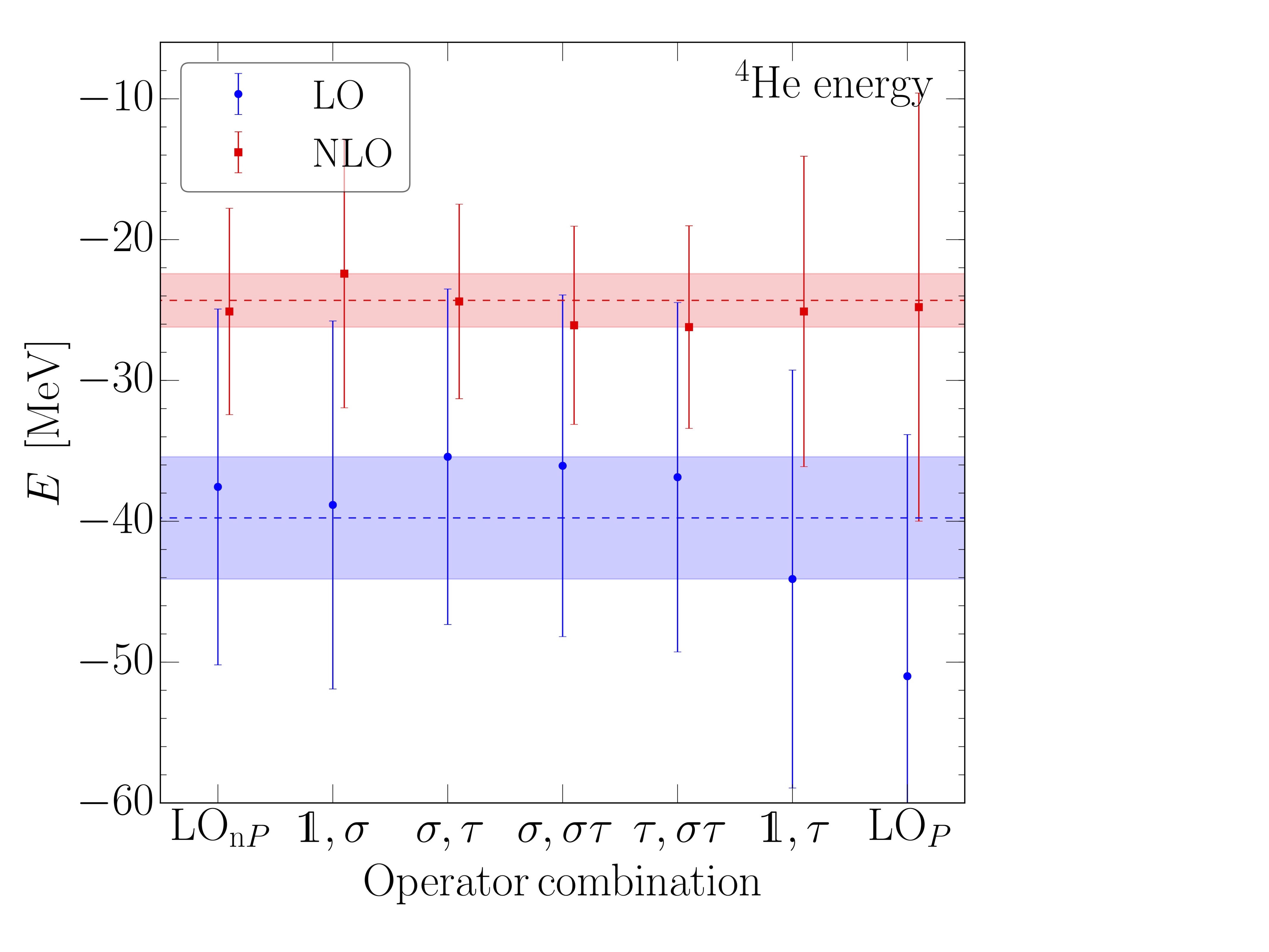}
\includegraphics[width=0.45\textwidth, trim=0 0.8cm 0 0, clip=]{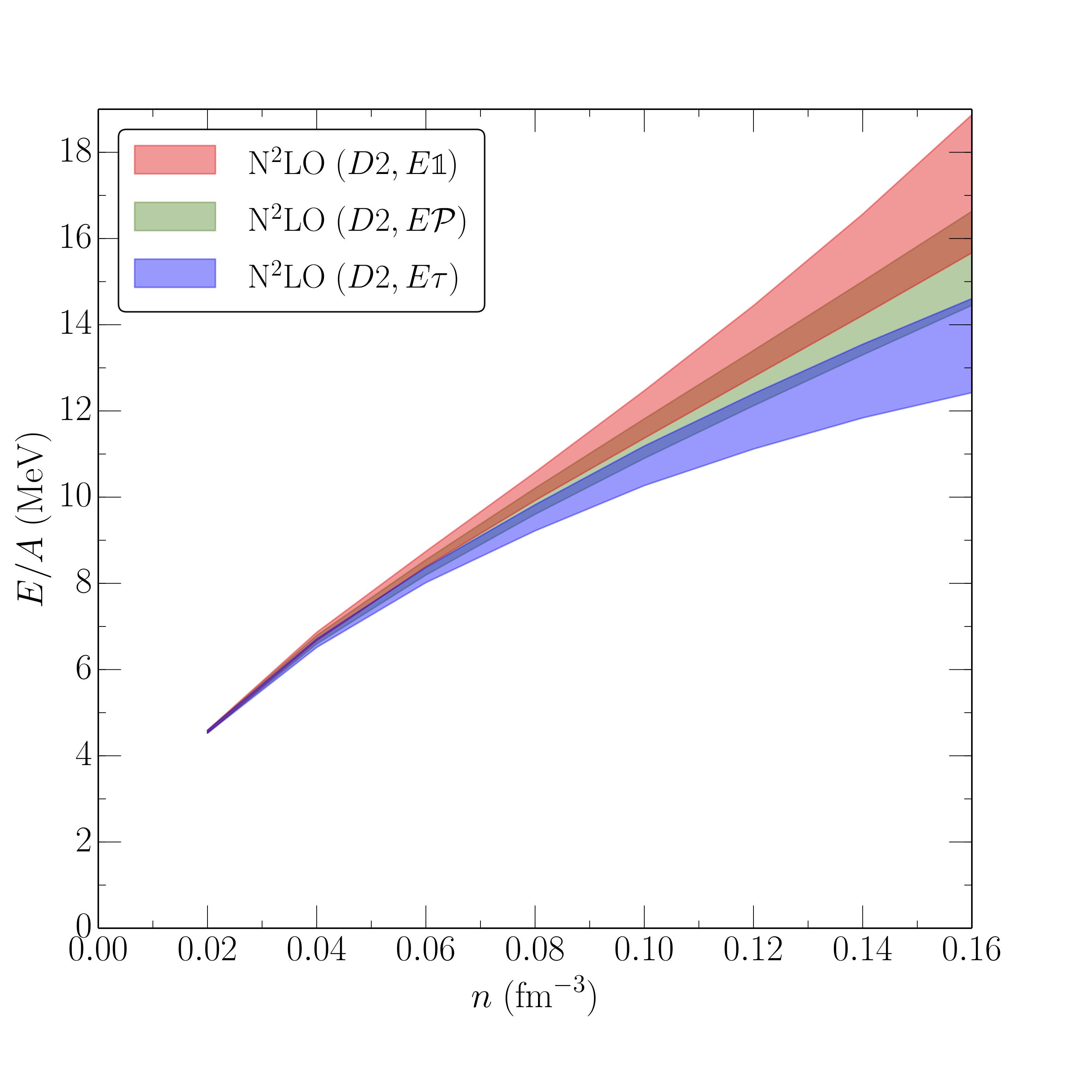}
\caption{\label{fig:Fierz}
Left: Ground-state energies of $^4$He at LO and NLO for different LO operator choices. Figure taken from Ref.~\cite{Huth:2017wzw} under the Creative Commons CCBY license. Right: Regulator artifacts in pure neutron matter due to the violation of the Fierz rearrangement freedom for the $3N$ contact interactions. The three different bands correspond to 3 different operator choices, where the green band projects the $3N$ interaction on triples with $S=1/2$ and $T=1/2$. Figure taken from Ref.~\cite{Lynn:2015jua} under the Creative Commons CCBY license.}
\end{figure} 

\subsection{Violation of Fierz-rearrangement freedom}~\label{sec:Fierz}

The first regulator artifact for local interactions affects short-range operators. 
In previous sections we had shown how only half of the operators at each order are linearly independent due to their insertion between antisymmetric fermionic states, see, e.g., Eq.~\eqref{eq:LO_antisymm}. However, this argument changes when a regulator function is applied. The discussion in this section will follow Ref.~\cite{Huth:2017wzw}.

In general, a regulator function can depend on two momentum scales, $f_{R}({\bf q},{\bf k})$. Local regulators, on the other hand, only depend on $\textbf{q}$, $f_{R, \textrm{loc}}({\bf q})$. 
The derivation of Eq.~\eqref{eq:LO_antisymm} remains valid if the regulator function commutes with the antisymmetrizer and, hence, reduces to a simple prefactor in Eq.~\eqref{eq:LO_antisymm}, i.e., when 
\begin{equation}
\label{eq:FierzCond}
f_R({\bf q},{\bf k})=f_R\left(-2{\bf k},-\frac12{\bf q}\right)\,.
\end{equation}
We can immediately see, that a purely local regulator can never fulfill this condition while typical nonlocal regulators of the form~\cite{Ekstrom:2015rta, Carlsson:2015vda,Entem:2017gor, Entem:2003ft,Epelbaum:2004fk}
\begin{equation} \label{eq:nonlocreg}
f_R({\bf p}, {\bf p}')=
\exp\left[-\left(\frac{{\bf p}\vphantom{'}}{\Lambda}\right)^{2n}\right]
\exp\left[-\left(\frac{{\bf p}'}{\Lambda}\right)^{2n}\right]\,,
\end{equation}
do. As a consequence,
\begin{equation}
\begin{split}
\label{eq:LO_antisymm_loc}
&V^{(0,\text{loc})}_{\text{cont,as}}=
\frac12\left(1-\frac{\mathcal{P}^{\text{m}}}{4}
(1+{\bm \sigma}_i \cdot {\bm \sigma}_j)(1+{\bm \tau}_i \cdot {\bm \tau}_j)\right)V_ {\text{cont}}^{(0)}f_R({\bf q})\\
&=\mathbbm{1}\left(\frac{C_{\mathbbm{1}}}{2}f_R({\bf q})-
\frac18\left(C_{\mathbbm{1}}+3C_{\sigma}+3C_{\tau}+
9C_{\sigma\tau}\right)f_R(2{\bf k})\right)\\
&+{\bm \sigma}_i \cdot {\bm \sigma}_j\left(\frac{C_{\sigma}}{2}f_R({\bf q})-
\frac18\left(C_{\mathbbm{1}}-C_{\sigma}+3C_{\tau}-
3C_{\sigma\tau}\right)f_R(2{\bf k})\right)\\
&+{\bm \tau}_i \cdot {\bm \tau}_j\left(\frac{C_{\tau}}{2}f_R({\bf q})-
\frac18\left(C_{\mathbbm{1}}+3C_{\sigma}-C_{\tau}-
3C_{\sigma\tau}\right)f_R(2{\bf k})\right)\\
&+{\bm \sigma}_i \cdot {\bm \sigma}_j{\bm \tau}_i \cdot {\bm \tau}_j\left(\frac{C_{\sigma\tau}}{2}f_R({\bf q})
-\frac18\left(C_{\mathbbm{1}}-C_{\sigma}-C_{\tau}+
C_{\sigma\tau}\right)f_R(2{\bf k})\right)\,,
\end{split}
\end{equation}
and the Fierz-rearrangement freedom is violated.  For general regulator functions as defined in the previous sections, this leads to 
\begin{equation}
V^{(0,\text{loc})}_{\text{cont,as}} = \left(\tilde C_S + \tilde C_T {\bm \sigma}_i \cdot {\bm \sigma}_j + \left(-\frac{2}{3}\tilde C_S-\tilde C_T \right) {\bm \tau}_i \cdot {\bm \tau}_j + \left(-\frac13 \tilde C_S\right) {\bm \sigma}_i \cdot {\bm \sigma}_j {\bm \tau}_i \cdot {\bm \tau}_j\right)f_R({\bf q})+ V_{\text{corr}}^f({\bf p \cdot p'})\,,
\end{equation}
where $V_\text{corr}^f({\bf p}\cdot{\bf p}')$ captures all the regulator artifacts that are of higher-order in the EFT. 
It depends on the functional form of the regulator and the cutoff value. 
One can also see, that the corrections can be angle-dependent, which leads to a mixing of different partial waves. 
As a consequence, when applying these regulators to a three-neutron system, for example, pure contact interactions, that otherwise would vanish due to the Pauli principle, start to contribute. 
This mixing of partial waves complicates the fitting procedure, increases theoretical uncertainties, and makes calculated observables dependent on the operator structure that was chosen. 

In Fig.~\ref{fig:Fierz} we show results for the $^4$He ground-state energy for different LO operator choices. As one can see, the ground-state energies can vary by approximately 10 MeV at LO, depending on the operator choice. However, when going to higher order and including subleading contact operators, regulator artifacts get partially absorbed and corrected. Then, only higher-order artifacts remain, which improves the situation considerably, as can be seen for the NLO results. In this case, the spread originating from different choices of LO operators reduces to approximately 4 MeV. 

A similar effect appears in the 3N sector, where the $V_E$ contact interaction suffers from a similar violation of the Fierz freedom when local regulators are applied. 
While $3N$ forces are typically fit to symmetric systems where this dependence can then be approximately accounted for, in triples with $S=3/2$ or $T=3/2$ (where typically no $3N$ contact force can contribute due to the Pauli principle) regulator artifacts appear, and lead to a finite contribution from $3N$ contact interactions that depend on the operator choice. 
We show this behavior in Fig.~\ref{fig:Fierz} in the right panel in the case of pure neutron matter, where all triples have T=3/2. 
The three different bands explore 3 choices for the $3N$ contact operators. 
At nuclear saturation density, we find that the regulator artifacts introduce a spread of approximately 5 MeV. Unfortunately, higher-order correction terms only appear at N$^4$LO and, to date, are not systematically included in any calculation.

Finally, we mention that the finite cutoff also introduces an ambiguity in the $V_D$ term, that depends on the choice of the initial spin-isospin structure when Fourier transforming:
\begin{align}
V_{D,1}^{ijk} &= \frac{g_A}{24f_{\pi}^2}\frac{c_D}{f_{\pi}^2 \Lambda_{\chi}}
\sum_{\pi(ijk)} { \bm \tau}_i\cdot { \bm \tau}_k \biggl[ \frac{m_{\pi}^3}{4 \pi} \, \delta(\textbf{r}_{ij}) X_{ik}(\textbf{r}_{kj}) - {\bm \sigma}_i \cdot {\bm \sigma}_k \, \delta(\textbf{r}_{ij}) \delta(\textbf{r}_{kj}) \biggr] \,, \\
V_{D,2}^{ijk} &= \frac{g_A}{24f_{\pi}^2}\frac{c_D}{f_{\pi}^2 \Lambda_{\chi}}
\sum_{\pi(ijk)} { \bm \tau}_i\cdot { \bm \tau}_k \biggl[ \frac{m_{\pi}^3}{4 \pi} \, \delta(\textbf{r}_{ij}) X_{ik}(\textbf{r}_{ik}) - {\bm \sigma}_i \cdot {\bm \sigma}_k \, \delta(\textbf{r}_{ij}) \delta(\textbf{r}_{ik}) \biggr] \,. \nonumber 
\end{align}
Both expressions are identical for true $\delta$ functions (infinite cutoff) but differ when a finite cutoff is applied. 

\subsection{Weaker pion exchanges}

A second regulator artifact for local regulators affects the pion exchanges. 
In Ref.~\cite{Tews:2015ufa} it was shown that locally regulated pion exchanges lead to less $3N$ repulsion than nonlocally regulated pion exchanges. 
At the Hartree-Fock level, for a typical cutoff of $2.5 \mathrm{fm}^{-1}$, when applying nonlocal regulators $\approx 97\%$ of the infinite cutoff result is recovered, while local regulators only recover $\approx60 \%$. To reproduce the momentum-space results, the cutoff has to be considerably increased. 

Local regulators for pion exchanges have been investigated in detail in Ref.~\cite{Dyhdalo:2016ygz} in both the $N\!N$ and $3N$ sector. The fact that the contribution due to pion exchanges is weaker for local than for nonlocal regulator functions is easy to understand in the Hartree-Fock approximation. 
At the Hartree-Fock level, there are both a direct and an exchange term. 
The momentum transfer $\textbf{q}=\textbf{p}-\textbf{p}'$ vanishes in the direct term because $\textbf{p}=\textbf{p}'$, but it is $\textbf{q}=2\textbf{p}$ in the exchange term because $\textbf{p}=-\textbf{p}'$. 
A typical local regulator of the form $\exp\left(-\left(\frac{q}{\Lambda} \right)^n \right)$, thus, evaluates to 1 in the direct term, but to $\exp\left(-\left(\frac{p}{\Lambda/2} \right)^n \right)$ in the exchange term. Therefore, compared to nonlocal regulators for which both terms are identical, $\exp\left(-\left(\frac{p}{\Lambda} \right)^n \right)$, local regulators have a very different behavior. In particular, local regulators have an effectively lower cutoff in the exchange channel. In the Hartree-Fock approximation, where the direct term vanishes for spin-dependent interactions like pion exchanges, only the exchange term contributes and, hence, is weaker for local than for nonlocal regulators. 

While the situation is more complicated when abandoning the Hartree-Fock approximation, this reasoning qualitatively remains valid and locally regulated pion exchanges are weaker than nonlocally regulated pion exchanges. 

\section{Selected results}
\label{sec:results}

In this section, we will briefly show the successes of Quantum Monte Carlo calculations with local chiral interactions for light atomic nuclei and infinite matter. 

\subsection{Light nuclei}

\begin{figure}[t]
\centering
\includegraphics[trim=0.0cm 0.5cm 0.1cm 1cm, clip=,width=0.53\textwidth,height=6cm]{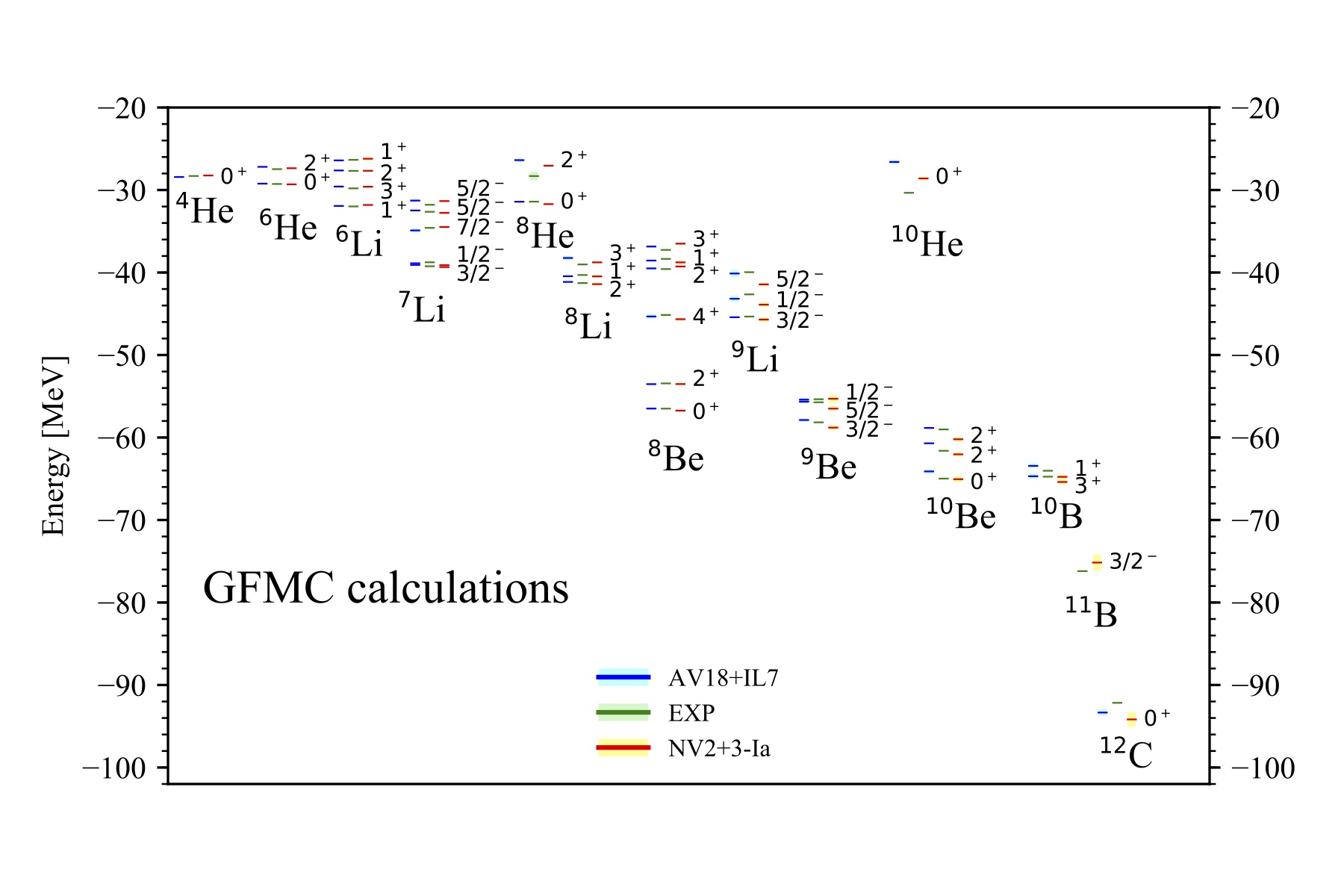}
\includegraphics[trim=0 0 0cm 0, clip=,width=0.42\textwidth,height=5.9cm]{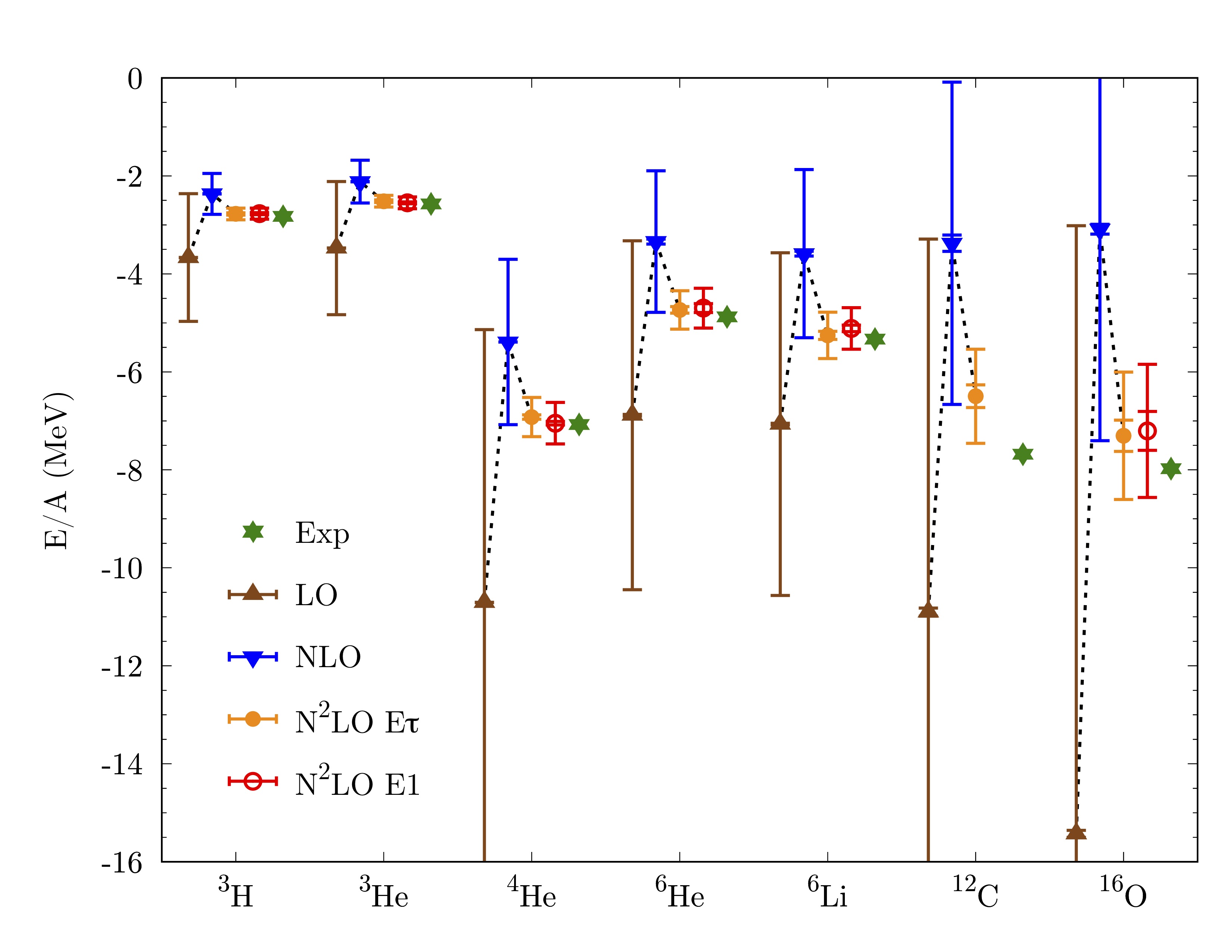}
\caption{\label{fig:NucleiResults}
Left: Spectra of light nuclei up to $^{12}$C obtained with GFMC with chiral interactions obtained in the delta-full approach (red) compared to experimental data (green) and GFMC calculations with phenomenological interactions (blue). Figure taken from Ref.~\cite{Piarulli:2017dwd} under the Creative Commons CCBY license. Right: Ground-state energies for nuclei up to $^{16}$O at different orders in the chiral expansion for AFDMC calculations with local interactions in the delta-less approach. Reprinted from Ref.~\cite{Lonardoni:2018nob} with permission from the American Physical Society.}
\end{figure} 

Local chiral interactions, both in the delta-less and delta-full approach, have been used to successfully describe properties of light nuclei using QMC methods. 
In Fig.~\ref{fig:NucleiResults}, we show GFMC results for ground- and excited states for nuclei up to $^{12}$C within the delta-full approach compared to experimental data. In addition, the results obtained with chiral EFT are compared to results with phenomenological interactions.
The results clearly show that chiral interactions describe spectra of light nuclei with great success and are compatible to the accuracy of phenomenological interaction in these systems. 
In addition, we also show ground-state energies obtained in the AFDMC method for nuclei up to $^{16}$O for delta-less chiral interactions. Results are given at LO, NLO, and N$^2$LO for two different $3N$ parameterizations to explore regulator artifacts. Again, chiral interactions agree well with experimental results, which are shown as green points.

\begin{figure}[t]
\centering
\includegraphics[trim=0 0 0cm 0, clip=,width=0.42\textwidth]{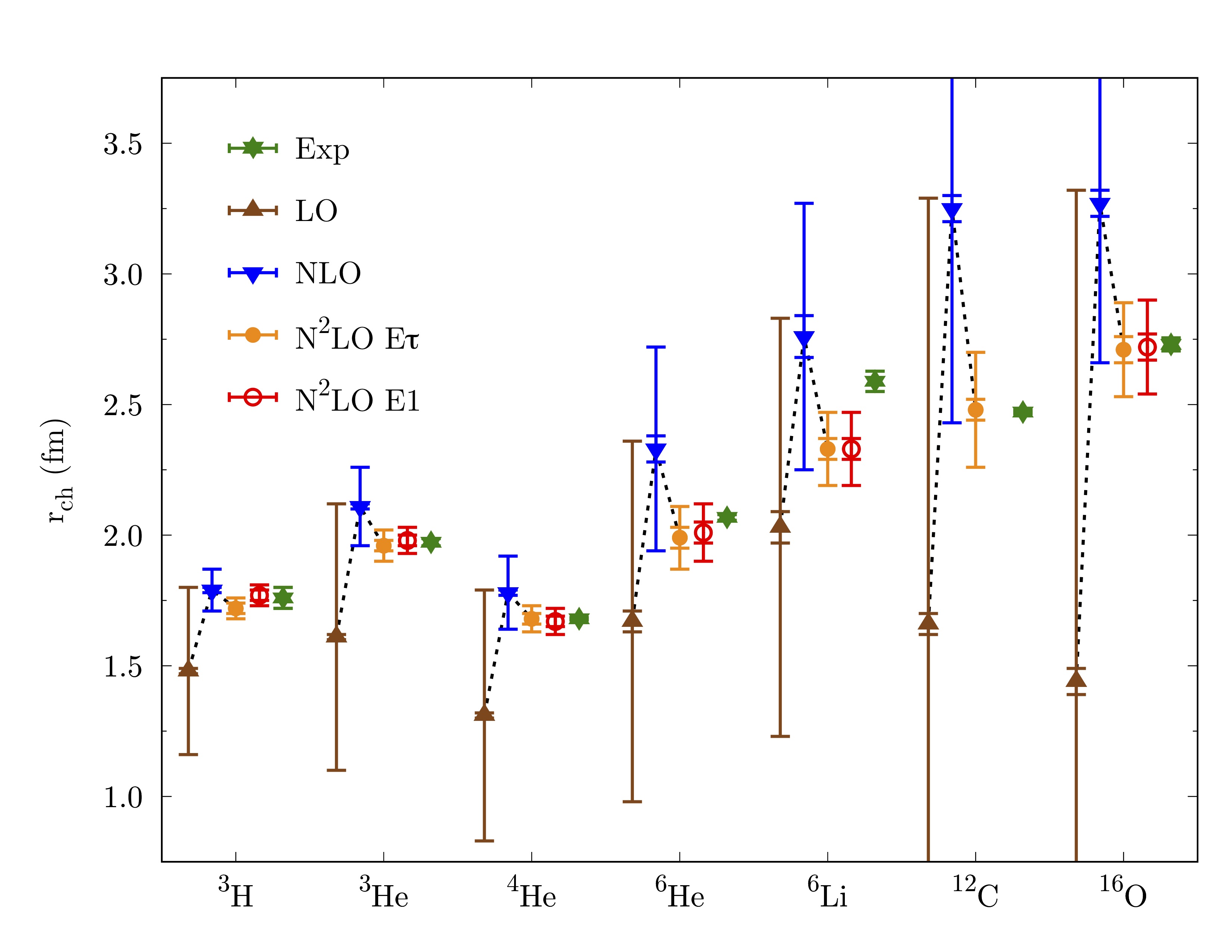}
\caption{\label{fig:ScattAndRadii}
Same as Fig.~\ref{fig:NucleiResults} (right) but for radii of nuclei up to $^{16}$O. Reprinted from Ref.~\cite{Lonardoni:2018nob} with permission from the American Physical Society.}
\end{figure} 

In addition to energies, local chiral interactions describe charge radii well. In Fig.~\ref{fig:ScattAndRadii}, we present order-by-order AFDMC results for the charge radii of nuclei up to $^{16}$O, compared to experiment. 
Again, the description is accurate. 
In addition, as mentioned before, delta-less chiral interactions have been adjusted to reproduce neutron-$\alpha$ scattering phase shifts, see Fig.~\ref{fig:3Nfits_Joel}. While $N\!N$ interactions alone cannot reproduce the $P$ wave splitting in this system (NLO calculations in Fig.~\ref{fig:3Nfits_Joel}),  chiral Hamiltonians at N$^2$LO, including $3N$ interactions, reproduce the neutron-$\alpha$ $P$ wave scattering phase shifts accurately. 

\subsection{Infinite Matter}

In addition to properties of atomic nuclei, local chiral interactions have been used to study infinite matter, and in particular, pure neutron matter.
In the right panel of Fig.~\ref{fig:Fierz}, we have already shown results for the energy per particle of pure neutron matter. 
Results are shown for 3 Hamiltonians at N$^2$LO, that explore the uncertainty due to regulator artifacts and the truncation of the chiral series. 
While uncertainties in pure neutron matter are enhanced due to the local regulator artifacts discussed before, indicated by the differences between the three bands, the resulting neutron-matter equation of state (EOS) is consistent with other ab initio determinations within uncertainties. 

These calculations have been successfully used to study the EOS of neutron stars, and it has been found that the resulting equations of state are consistent with astrophysical observations of pulsar masses. The EOS have also been used to study gravitational waves from neutron-star mergers~\cite{Tews:2018chv,Tews:2019cap,Capano:2019eae}. 

\section{Conclusion and outlook}
The quest to understand properties of nuclear systems in terms of forces acting between the nucleons has been considered one of the most challenging efforts of nuclear theory. During the past quarter century, particular emphasis has been devoted to the systematic framework provided by chiral EFT. This approach allows for a consistent description of the two- and many-body interactions and ensuing many-body currents, and a quantification of the theoretical uncertainty due to the truncation error in the chiral expansion.

In this review, we have presented a comprehensive description of the two families of local chiral interactions that have been developed for the use in QMC methods: one within the delta-less and one within the delta-full approach. We provided many details about the theoretical derivation and optimization of these nuclear models addressing their similarities and differences. For completeness, we also presented selected QMC results for light nuclei and neutron matter. These results show that the combination of local chiral EFT interactions with powerful QMC many-body methods can accurately describe ground- and excited-state energies, radii of nuclei up to $^{16}$O, and $n$-$\alpha$ scattering, as well as the equation of state of neutron matter. 

These local chiral interactions have also been used to calculate the distribution of nucleons in a nucleus in both momentum and coordinate space which are related to experimental observations~\cite{Lonardoni:2018nob,Lonardoni:2018sqo,Cruz-Torres:2019fum,Lynn:2019vwp}, in benchmark calculations of the energy per particle of pure neutron matter as a function of the baryon density~\cite{Piarulli:2019pfq} and in studies of neutrinoless double-beta decays~\cite{Cirigliano:2019vdj}.

In future, local chiral interactions will continue to serve as input for precise QMC methods to systematically study, for example, electroweak reactions, along the lines of Refs.~\cite{Pastore:2012rp,Marcucci:2015rca,Lovato:2016gkq,Lovato:2017cux,Pastore:2017uwc,Schiavilla:2018udt,Pastore:2019urn} and infinite matter also at finite proton fractions.

Improvements to the interactions that reduce uncertainties due to the scheme and scale dependence of the interactions, e.g., the inclusion of higher orders in the chiral expansion in both the $NN$ and $3N$ sectors, will provide exciting prospects and permit precision studies of many nuclear systems.

%In addition, this combination has been found to also accurately describe electroweak reactions. The fact that these different results are in excellent agreement with experimental knowledge highlights the promise that lies in this approach. 

%%%%%%%%%%%%%%%%%%%%%%%%%%%%%%5

\section*{Conflict of Interest Statement}
%All financial, commercial or other relationships that might be perceived by the academic community as representing a potential conflict of interest must be disclosed. If no such relationship exists, authors will be asked to confirm the following statement: 

The authors declare that the research was conducted in the absence of any commercial or financial relationships that could be construed as a potential conflict of interest.

\section*{Author Contributions}

Both authors contributed to this review in equal parts.

\section*{Funding}
This work was supported by the US Department of Energy, Office of Science, Office of Nuclear Physics, under Contract DE-AC52-06NA25396 (IT), the FRIB Theory Alliance award DE-SC0013617 (MP), the Los Alamos National Laboratory (LANL) LDRD program, and the NUCLEI SciDAC and INCITE programs. Part of this research used resources of the Argonne Leadership Computing Facility at Argonne National Laboratory, the Los Alamos Open Supercomputing via the Institutional Computing (IC) program, and the National Energy Research Scientific Computing Center (NERSC), which is supported by the U.S. Department of Energy, Office of Science, under contract No.~DE-AC02-05CH11231.

\section*{Acknowledgments}
We thank our collaborators A. Baroni, J. Carlson, E. Epelbaum, S. Gandolfi, A. Gezerlis, L. Girlanda, K. Hebeler, L. Huth, A. Kievsky, D. Lonardoni, A. Lovato, J. E. Lynn, L. E. Marcucci, A. Nogga, S. Pastore, R. Schiavilla, K. E. Schmidt, A. Schwenk, M. Viviani and R. B. Wiringa for their contributions to the studies presented in this work.

\appendix

\section*{Appendix A: Relevant expressions for the coordinate-space representation of TPE and contact interactions}
\label{supl:s1}

In the following appendix, we list the relevant expressions for the coordinate-space representation of TPE and contact interaction radial functions for the delta-full approach introduced in Section 3.
\subsection*{Two-pion-exchange}
As stated in Section 3, the TPE interaction can be specified in terms of the following radial functions:
\begin{eqnarray}
V_{C,\pi}(r)&=&v^{2\pi,\rm NLO}_{c}(r;1\Delta)
+v^{2\pi,\rm NLO}_{c}(r;2\Delta)+v^{2\pi,\rm N2LO}_{c}(r;\slashed{\Delta})+v^{2\pi,\rm N2LO}_{c}(r;1\Delta)\nonumber\\
&&+v^{2\pi,\rm N2LO}_{c}(r;2\Delta)\ , \label{eq:cL}\\
V_S(r)&=&v^{2\pi,\rm NLO}_\sigma (r;\slashed{\Delta})
+v^{2\pi,\rm NLO}_{\sigma}(r;1\Delta)+v^{2\pi,\rm NLO}_{\sigma}(r;2\Delta)+v^{2\pi,\rm N2LO}_{\sigma}(r;1\Delta)\nonumber\\
&&+v^{2\pi,\rm N2LO}_{\sigma}(r;2\Delta) \ , \\
V_T(r)&=&v^{2\pi,\rm NLO}_t (r;\slashed{\Delta})
+v^{2\pi,\rm NLO}_{t}(r;1\Delta)+v^{2\pi,\rm NLO}_{t}(r;2\Delta)+v^{2\pi,\rm N2LO}_{t}(r;1\Delta)\nonumber\\
&&+v^{2\pi,\rm N2LO}_{t}(r;2\Delta) \ , 
\end{eqnarray}
\begin{eqnarray}
W_C(r)&=&v^{2\pi,\rm NLO}_\tau (r;\slashed{\Delta})
+v^{2\pi,\rm NLO}_{\tau}(r;1\Delta)+v^{2\pi,\rm NLO}_{\tau}(r;2\Delta)+v^{2\pi,\rm N2LO}_{\tau}(r;1\Delta)\nonumber\\
&&+v^{2\pi,\rm N2LO}_{\tau}(r;2\Delta) \ , \\
W_S(r)&=&v^{\pi,\rm LO}_{\sigma\tau}(r)+v^{2\pi,\rm NLO}_{\sigma\tau}(r;1\Delta)
+v^{2\pi,\rm NLO}_{\sigma\tau}(r;2\Delta)+v^{2\pi,\rm N2LO}_{\sigma\tau}(r;\slashed{\Delta})
\nonumber\\
&&+v^{2\pi,\rm N2LO}_{\sigma\tau}(r;1\Delta)+v^{2\pi,\rm N2LO}_{\sigma\tau}(r;2\Delta)\ ,\\
W_T(r)&=&v^{\pi,\rm LO}_{t\tau}(r)+
v^{2\pi,\rm NLO}_{t\tau}(r;1\Delta)
+v^{2\pi,\rm NLO}_{t\tau}(r;2\Delta)+v^{2\pi,\rm N2LO}_{t\tau}(r;\slashed{\Delta})
\nonumber\\
&&+v^{2\pi,\rm N2LO}_{t\tau}(r;1\Delta)+v^{2\pi,\rm N2LO}_{t\tau}(r;2\Delta)\, ,\label{eq:ttauL}
\end{eqnarray}
where each of these functions is multiplied by the cutoff $f_{\rm long}^{\Delta}(r)$ defined in Section 3.1.2.\\
%\begin{equation}
%v^l_{\rm L}(r) \longrightarrow C_{R_{\rm L}}(r) \, v^l_{\rm L}(r) \ ,
%\end{equation}
%with $l=c,\tau,\sigma,\sigma\tau,t,t\tau,\sigma T,tT$.
\indent Below, we give the coordinate-space representation of the TPE radial functions at NLO and N2LO that contribute to $V_{C,\pi}$, $W_C$, $V_S$, $W_S$, $V_T$, and $W_T$.\\
%appearing in Eqs.~(\ref{eq:Lr_1})-(\ref{eq:Lr_2}).
%The NLO terms corresponding to diagrams (d)--(f) read as
\indent At NLO, the terms without $\Delta$'s read as
\begin{eqnarray}
v^{2\pi,\rm NLO}_\sigma (r;\slashed{\Delta})&=&\frac{1}
{2 \pi^3  r^4}\frac{g_A^4}{(2f_{\pi})^4}\,m_{\pi} \bigg[3x\,K_0(2x)+
(3+2x^2)K_1(2x)\bigg]\ , \label{eq:NLOr1}\\
v^{2\pi,\rm NLO}_t (r;\slashed{\Delta})&=&-\frac{1}
{8 \pi^3 r^4}\frac{g_A^4}{(2f_{\pi})^4 }\,m_{\pi} \bigg[12x\,K_0(2x)+
(15+4x^2)K_1(2x)\bigg] \ , \label{eq:NLOr2}\\
v^{2\pi,\rm NLO}_\tau (r;\slashed{\Delta})&=&\frac{1}
{8 \pi^3  r^4}\frac{m_\pi}{(2f_{\pi})^4}\, \bigg[\, x
\left[1+10 g_A^2-g_A^4 (23+4x^2)\right]K_0(2x)\nonumber\\
&&+\left[1+2 g_A^2 (5+2x^2)-g_A^4 (23+12x^2)\right]K_1(2x)\bigg] \ ,\label{eq:NLOr3}
\end{eqnarray}
where $g_A$, $f_\pi$, and $m_\pi$ denote the axial-vector coupling constant of the nucleon, the pion decay constant, and the pion mass, respectively, $x=m_\pi r$ and $K_n$ are modified Bessel functions of the second kind.\\
\indent The NLO terms with a single $\Delta$ intermediate state are given by
\begin{eqnarray}
v^{2\pi,\rm NLO}_{c} (r;1\Delta)&=&-\frac{1}
{6 \pi^2  r^5\,y}\frac{g_A^2 h_A^2}{(2f_{\pi})^4}e^{-2x}\left(6+12x+10x^2+4x^3+x^4\right)\ , \label{eq:NLOdr1}\\
v^{2\pi,\rm NLO}_{\sigma}(r;1\Delta)&=&-\frac{1}
{72 \pi^3  r^5}\frac{g_A^2 h_A^2}{(2f_{\pi})^4}\Bigg[2\,\int_{0}^{\infty}d\mu \frac{\mu^2}
{\sqrt{\mu^2+4x^2}}e^{-\sqrt{\mu^2+4x^2}}(\mu^2+4x^2)\nonumber \\
&&-\frac{1}{y}\int_{0}^{\infty}d\mu \frac{\mu}
{\sqrt{\mu^2+4x^2}}e^{-\sqrt{\mu^2+4x^2}}(\mu^2+4x^2)\nonumber\\
&&\times(\mu^2+4y^2)\arctan{\frac{\mu}{2y}}\Bigg]\ ,\label{eq:NLOdr2}\\
v^{2\pi,\rm NLO}_{t}(r;1\Delta)&=&\frac{1}
{144 \pi^3  r^5}\frac{g_A^2 h_A^2}{(2f_{\pi})^4}\Bigg[2\,\int_{0}^{\infty}d\mu \frac{\mu^2}
{\sqrt{\mu^2+4x^2}}e^{-\sqrt{\mu^2+4x^2}}(3+3\sqrt{\mu^2+4x^2}\nonumber \\
&&+\mu^2+4x^2)-\frac{1}{y}\int_{0}^{\infty}d\mu\frac{\mu}
{\sqrt{\mu^2+4x^2}}e^{-\sqrt{\mu^2+4x^2}}(\mu^2+4y^2)\nonumber\\
&&\times(3+3\sqrt{\mu^2+4x^2}
+\mu^2+4x^2)\arctan{\frac{\mu}{2y}}\Bigg]\ , \label{eq:NLOdr3}
\end{eqnarray}
\begin{eqnarray}
v^{2\pi,\rm NLO}_{\tau} (r;1\Delta)&=&-\frac{1}
{216 \pi^3  r^5}\frac{h_A^2}{(2f_{\pi})^4}\Bigg[\int_{0}^{\infty}d\mu \frac{\mu^2}
{\sqrt{\mu^2+4x^2}}e^{-\sqrt{\mu^2+4x^2}}(12x^2+5\mu^2+12y^2)\nonumber\\
&&-12y\int_{0}^{\infty}d\mu \frac{\mu}
{\sqrt{\mu^2+4x^2}}e^{-\sqrt{\mu^2+4x^2}}(2x^2+\mu^2+2y^2)\arctan{\frac{\mu}{2y}}\Bigg]\nonumber\\
&&-\frac{1}{216 \pi^3  r^5}\frac{g_A^2 h_A^2}{(2f_{\pi})^4}\Bigg[-\int_{0}^{\infty}d\mu \frac{\mu^2}
{\sqrt{\mu^2+4x^2}}e^{-\sqrt{\mu^2+4x^2}}\nonumber\\
&&\times(24x^2+11\mu^2+12y^2)+\frac{6}{y}\int_{0}^{\infty}d\mu \frac{\mu}
{\sqrt{\mu^2+4x^2}}e^{-\sqrt{\mu^2+4x^2}}\nonumber\\
&&\times(2x^2+\mu^2+2y^2)^2\arctan{\frac{\mu}{2y}}\Bigg]\ ,\label{eq:NLOdr4}\\
v^{2\pi,\rm NLO}_{\sigma\tau} (r;1\Delta)&=&\frac{1}
{54 \pi^2  r^5\,y}\frac{g_A^2 h_A^2}{(2f_{\pi})^4}e^{-2x}\left(1+x \right)\left(3+3x+x^2\right)\ , \label{eq:NLOdr5}\\
v^{2\pi,\rm NLO}_{t\tau} (r;1\Delta)&=&-\frac{1}
{54 \pi^2  r^5\,y}\frac{g_A^2 h_A^2}{(2f_{\pi})^4}e^{-2x}\left(1+x \right)\left(3+3x+2x^2\right)\ ,\label{eq:NLOdr6}
\end{eqnarray}
where $h_A$ is the $N$-to-$\Delta$ axial coupling
constant, $y= \Delta r$ ($\Delta$ is the $\Delta$-nucleon mass difference) and the parametric integral over $\mu$ is carried out numerically. \\
\indent The NLO terms with $2\,\Delta$ intermediate states are
\begin{eqnarray}
v^{2\pi,\rm NLO}_{c} (r;2\Delta)&=&-\frac{1}
{108 \pi^3  r^5}\frac{h_A^4}{(2f_{\pi})^4}\Bigg[\int_{0}^{\infty}d\mu\frac{\mu^2}
{\sqrt{\mu^2+4x^2}}e^{-\sqrt{\mu^2+4x^2}}
\bigg[4y^2\nonumber\\
&&+2\frac{(2x^2+\mu^2+2y^2)^2}{(\mu^2+4y^2)}\bigg]+\frac{1}{y}\int_{0}^{\infty}d\mu\frac{\mu}
{\sqrt{\mu^2+4x^2}}e^{-\sqrt{\mu^2+4x^2}}\nonumber\\
&&\times(2x^2+\mu^2+2y^2)(2x^2+\mu^2-6y^2)\arctan{\frac{\mu}{2y}}\Bigg]\ ,\label{eq:NLO2dr1}\\
v^{2\pi,\rm NLO}_{\sigma} (r;2\Delta)&=&-\frac{1}
{1296 \pi^3  r^5}\frac{h_A^4}{(2f_{\pi})^4}\Bigg[-6 \int_{0}^{\infty}d\mu\frac{\mu^2}
{\sqrt{\mu^2+4x^2}}e^{-\sqrt{\mu^2+4x^2}}
(\mu^2+4x^2)\nonumber\\
&&+\frac{1}{y}\int_{0}^{\infty}d\mu\frac{\mu}
{\sqrt{\mu^2+4x^2}}e^{-\sqrt{\mu^2+4x^2}}(\mu^2+4x^2)
(\mu^2+12y^2)\nonumber\\
&&\times\arctan{\frac{\mu}{2y}}\Bigg]\ ,\label{eq:NLO2dr2}
\end{eqnarray}
\begin{eqnarray}
v^{2\pi,\rm NLO}_{t} (r;2\Delta)&=&\frac{1}
{2592 \pi^3  r^5}\frac{h_A^4}{(2f_{\pi})^4}\Bigg[-6 \int_{0}^{\infty}d\mu\frac{\mu^2}
{\sqrt{\mu^2+4x^2}}e^{-\sqrt{\mu^2+4x^2}}
(3+3\sqrt{\mu^2+4x^2}\nonumber\\
&&+\mu^2+4x^2)+\frac{1}{y}\int_{0}^{\infty}d\mu\frac{\mu}
{\sqrt{\mu^2+4x^2}}e^{-\sqrt{\mu^2+4x^2}}(3+3\sqrt{\mu^2+4x^2}\nonumber\\
&&+\mu^2+4x^2)
(\mu^2+12y^2)\arctan{\frac{\mu}{2y}}\Bigg]\ ,\label{eq:NLO2dr3}\\
v^{2\pi,\rm NLO}_{\tau} (r;2\Delta)&=&-\frac{1}
{1944 \pi^3  r^5}\frac{h_A^4}{(2f_{\pi})^4}\Bigg[\int_{0}^{\infty}d\mu\frac{\mu^2}
{\sqrt{\mu^2+4x^2}}e^{-\sqrt{\mu^2+4x^2}}\bigg[(24x^2+11\mu^2\nonumber\\
&&+24y^2)+6\frac{(2x^2+\mu^2+2y^2)^2}{(\mu^2+4y^2)}\bigg]-\frac{3}{y}\int_{0}^{\infty}d\mu\frac{\mu}
{\sqrt{\mu^2+4x^2}}e^{-\sqrt{\mu^2+4x^2}}\nonumber\\
&&\times(2x^2+\mu^2+2y^2)(2x^2+\mu^2+10y^2)\arctan{\frac{\mu}{2y}}\Bigg]\ ,\label{eq:NLO2dr4}\\
v^{2\pi,\rm NLO}_{\sigma\tau} (r;2\Delta)&=&-\frac{1}
{7776 \pi^3  r^5}\frac{h_A^4}{(2f_{\pi})^4}\Bigg[-2 \int_{0}^{\infty}d\mu\frac{\mu^2}
{\sqrt{\mu^2+4x^2}}e^{-\sqrt{\mu^2+4x^2}}
(\mu^2+4x^2)\nonumber\\
&&+\frac{1}{y}\int_{0}^{\infty}d\mu\frac{\mu}
{\sqrt{\mu^2+4x^2}}e^{-\sqrt{\mu^2+4x^2}}(\mu^2+4x^2)
(-\mu^2+4y^2)\nonumber\\
&&\times\arctan{\frac{\mu}{2y}}\Bigg]\ ,\label{eq:NLO2dr5}
\end{eqnarray}
\begin{eqnarray}
v^{2\pi,\rm NLO}_{t\tau} (r;2\Delta)&=&\frac{1}
{15552 \pi^3  r^5}\frac{h_A^4}{(2f_{\pi})^4}\Bigg[-2 \int_{0}^{\infty}d\mu\frac{\mu^2}
{\sqrt{\mu^2+4x^2}}e^{-\sqrt{\mu^2+4x^2}}
(3+3\sqrt{\mu^2+4x^2}\nonumber\\
&&+\mu^2+4x^2)+\frac{1}{y}\int_{0}^{\infty}d\mu\frac{\mu}
{\sqrt{\mu^2+4x^2}}e^{-\sqrt{\mu^2+4x^2}}(3+3\sqrt{\mu^2+4x^2}\nonumber\\
&&+\mu^2+4x^2)
(-\mu^2+4y^2)\arctan{\frac{\mu}{2y}}\Bigg]\ .\label{eq:NLO2dr6}
\end{eqnarray}

Moving on to the loop corrections at N2LO, the functions are given by
\begin{eqnarray}
v^{2\pi,\rm N2LO}_c (r;\slashed{\Delta})&=&\frac{3}{2\,\pi^2 r^6}\frac{g_A^2}{(2f_{\pi})^4}e^{-2x}
[2c_1x^2(1+x)^2+c_3(6+12x+10x^2\nonumber\\
&&+4x^3+x^4)]\ , \label{eq:N2LOr1}\\
v^{2\pi,\rm N2LO}_{\sigma\tau} (r;\slashed{\Delta})&=&\frac{1}{3\,\pi^2 r^6}\frac{g_A^2}{(2f_{\pi})^4}c_4 e^{-2x}
\left(1+x\right)\left(3+3x+2x^2\right)\ , \label{eq:N2LOr2}\\
v^{2\pi,\rm N2LO}_{t\tau} (r;\slashed{\Delta})&=&-\frac{1}{3\,\pi^2 r^6}\frac{g_A^2}{(2f_{\pi})^4}c_4 e^{-2x}
\left(1+x\right)\left(3+3x+x^2\right)\ , \label{eq:N2LOr3}
\end{eqnarray}

while those with 1 $\Delta$ are given by
\begin{eqnarray}
v^{2\pi,\rm N2LO}_{c} (r;1\Delta)&=&\frac{1}
{18 \pi^3  r^6}\frac{h_A^2\,y}{(2f_{\pi})^4}\Bigg[\int_{0}^{\infty}d\mu\frac{\mu^2}
{\sqrt{\mu^2+4x^2}}e^{-\sqrt{\mu^2+4x^2}}
[-24c_1x^2\nonumber\\
&&+c_2(5\mu^2+12x^2+12y^2)-6c_3(\mu^2+2x^2)]\nonumber\\
&&+\frac{6}{y}\int_{0}^{\infty}\!\!d\mu\frac{\mu}
{\sqrt{\mu^2+4x^2}} e^{-\sqrt{\mu^2+4x^2}}(\mu^2\!+\!2x^2\!+\!2y^2) 
\nonumber\\
&&\times[4c_1x^2-2c_2y^2+c_3(\mu^2+2x^2)]\,\arctan{\frac{\mu}{2y}}\Bigg]\ ,\label{eq:N2LOdr1}\\
v^{2\pi,\rm N2LO}_{\sigma} (r;1\Delta)&=&-\frac{1}
{18 \pi^3  r^6}\frac{(b_3+b_8)\,h_A\,g_A^2\,y}{(2f_{\pi})^4}\Bigg[2\int_{0}^{\infty}d\mu\frac{\mu^2}
{\sqrt{\mu^2+4x^2}}e^{-\sqrt{\mu^2+4x^2}}\nonumber\\
&&\times (\mu^2+4x^2)-\frac{1}{y}\int_{0}^{\infty}d\mu\frac{\mu}{\sqrt{\mu^2+4x^2}}e^{-\sqrt{\mu^2+4x^2}}
(\mu^2+4x^2)\nonumber\\
&&\times(\mu^2+4y^2)\,\arctan{\frac{\mu}{2y}}\Bigg]\ ,\label{eq:N2LOdr2}\\
v^{2\pi,\rm N2LO}_{t} (r;1\Delta)&=&\frac{1}
{36 \pi^3  r^6}\frac{(b_3+b_8)\,h_A\,g_A^2\,y}{(2f_{\pi})^4}\Bigg[2\int_{0}^{\infty}d\mu\frac{\mu^2}
{\sqrt{\mu^2+4x^2}}e^{-\sqrt{\mu^2+4x^2}}\nonumber\\
&&(3+3\sqrt{\mu^2+4x^2}+\mu^2+4x^2)-\frac{1}{y}\int_{0}^{\infty}d\mu\frac{\mu}{\sqrt{\mu^2+4x^2}}\nonumber\\
&&e^{-\sqrt{\mu^2+4x^2}}(3+3\sqrt{\mu^2+4x^2}+\mu^2+4x^2)\nonumber\\
&&\times(\mu^2+4y^2)\arctan{\frac{\mu}{2y}}\Bigg]\ ,\label{eq:N3LOdr1}\\
v^{2\pi,\rm N2LO}_{\tau} (r;1\Delta)&=&-\frac{1}
{54 \pi^3  r^6}\frac{(b_3+b_8)\,h_A\,y}{(2f_{\pi})^4}\Bigg[+\int_{0}^{\infty}d\mu\frac{\mu^2}
{\sqrt{\mu^2+4x^2}}e^{-\sqrt{\mu^2+4x^2}}\nonumber\\
&&\times(5\mu^2+12x^2+12y^2)-12\,y\int_{0}^{\infty}d\mu\frac{\mu}
{\sqrt{\mu^2+4x^2}}e^{-\sqrt{\mu^2+4x^2}}\nonumber\\
&&\times(\mu^2+2x^2+2y^2)\,\arctan{\frac{\mu}{2y}}\Bigg]-\frac{1}{54 \pi^3  r^6}\frac{(b_3+b_8)\,h_A\,g_A^2\,y}{(2f_{\pi})^4}\nonumber\\
&&\times\Bigg[-\int_{0}^{\infty}d\mu\frac{\mu^2}
{\sqrt{\mu^2+4x^2}}e^{-\sqrt{\mu^2+4x^2}}(11\mu^2+24x^2+12y^2)\nonumber\\
&&+\frac{6}{y}\int_{0}^{\infty}d\mu\frac{\mu}{\sqrt{\mu^2+4x^2}}e^{-\sqrt{\mu^2+4x^2}}
\left(\mu^2+2x^2+2y^2\right)^2\nonumber\\
&&\times\arctan{\frac{\mu}{2y}}\Bigg]\ ,\label{eq:N2LOdr4}
\end{eqnarray}
\begin{eqnarray}
v^{2\pi,\rm N2LO}_{\sigma\tau} (r;1\Delta)&=&-\frac{1}
{108 \pi^3  r^6}\frac{c_4\,h_A^2\,y}{(2f_{\pi})^4}\Bigg[2\,\int_{0}^{\infty}d\mu\frac{\mu^2}
{\sqrt{\mu^2+4x^2}}e^{-\sqrt{\mu^2+4x^2}}(\mu^2+4x^2)\nonumber\\
&&-\frac{1}{y}\int_{0}^{\infty}d\mu\frac{\mu}
{\sqrt{\mu^2+4x^2}}e^{-\sqrt{\mu^2+4x^2}}(\mu^2+4x^2)(\mu^2+4y^2)\nonumber\\
&&\times\arctan{\frac{\mu}{2y}}\Bigg]\ ,\label{eq:N2LOdr5}\\
v^{2\pi,\rm N2LO}_{t\tau} (r;1\Delta)&=&\frac{1}
{216 \pi^3  r^6}\frac{c_4\,h_A^2\,y}{(2f_{\pi})^4}\Bigg[2\,\int_{0}^{\infty}d\mu\frac{\mu^2}
{\sqrt{\mu^2+4x^2}}e^{-\sqrt{\mu^2+4x^2}}(3+3\sqrt{\mu^2+4x^2}\nonumber\\
&&+\mu^2+4x^2)-\frac{1}{y}\int_{0}^{\infty}d\mu\frac{\mu}
{\sqrt{\mu^2+4x^2}}e^{-\sqrt{\mu^2+4x^2}}(3+3\sqrt{\mu^2+4x^2}\nonumber\\
&&+\mu^2+4x^2)(\mu^2+4y^2)\,\arctan{\frac{\mu}{2y}}\Bigg]\  .\label{eq:N2LOdr6}
\end{eqnarray}
\indent Lastly, the contributions with 2$\Delta$ read
\begin{eqnarray}
v^{2\pi,\rm N2LO}_{c} (r;2\Delta)&=&-\frac{2}
{81 \pi^3  r^6}\frac{(b_3+b_8)\,h_A^3\,y}{(2f_{\pi})^4}\Bigg[
\int_{0}^{\infty}d\mu\frac{\mu^2}{\sqrt{\mu^2+4x^2}}e^{-\sqrt{\mu^2+4x^2}}\nonumber\\
&&\times[6\frac{(\mu^2+2x^2+2y^2)^2}{\mu^2+4y^2}+11\mu^2+24x^2+12y^2]\nonumber\\
&&-\frac{3}{y}\int_{0}^{\infty}d\mu\frac{\mu}
{\sqrt{\mu^2+4x^2}}e^{-\sqrt{\mu^2+4x^2}}(\mu^2+2x^2+10y^2)\nonumber\\
&&\times(\mu^2+2x^2+2y^2)
\arctan{\frac{\mu}{2y}}\Bigg]\ ,  \label{eq:N2LO2dr1}\\
v^{2\pi,\rm N2LO}_{\sigma\tau} (r;2\Delta)&=&-\frac{1}
{972 \pi^3  r^6}\frac{(b_3+b_8)\,h_A^3\,y}{(2f_{\pi})^4}\Bigg[
-6\int_{0}^{\infty}d\mu\frac{\mu^2}{\sqrt{\mu^2+4x^2}}e^{-\sqrt{\mu^2+4x^2}}
\nonumber\\
&&\times(\mu^2+4x^2)+\frac{1}{y}\int_{0}^{\infty}d\mu\frac{\mu}
{\sqrt{\mu^2+4x^2}}e^{-\sqrt{\mu^2+4x^2}}(\mu^2+4x^2)\nonumber\\
&&\times(\mu^2+12y^2)\arctan{\frac{\mu}{2y}}\Bigg]\ , \label{eq:N2LO2dr2}
\end{eqnarray}
\begin{eqnarray}
v^{2\pi,\rm N2LO}_{t} (r;2\Delta)&=&\frac{1}
{324 \pi^3  r^6}\frac{(b_3+b_8)\,h_A^3\,y}{(2f_{\pi})^4}\Bigg[
-6\int_{0}^{\infty}d\mu\frac{\mu^2}{\sqrt{\mu^2+4x^2}}e^{-\sqrt{\mu^2+4x^2}}\nonumber\\
&&\times(3+3\sqrt{\mu^2+4x^2}+\mu^2+4x^2)+\frac{1}{y}\int_{0}^{\infty}d\mu\frac{\mu}
{\sqrt{\mu^2+4x^2}}e^{-\sqrt{\mu^2+4x^2}}\nonumber\\
&&\times(3+3\sqrt{\mu^2+4x^2}+\mu^2+4x^2)(\mu^2+12y^2)
\arctan{\frac{\mu}{2y}}\Bigg]\ ,  \label{eq:N2LO2dr3}\\
v^{2\pi,\rm N2LO}_{\tau} (r;2\Delta)&=&-\frac{1}
{243 \pi^3  r^6}\frac{(b_3+b_8)\,h_A^3\,y}{(2f_{\pi})^4}\Bigg[
\int_{0}^{\infty}d\mu\frac{\mu^2}{\sqrt{\mu^2+4x^2}}e^{-\sqrt{\mu^2+4x^2}}\nonumber\\
&&\times[6\frac{(\mu^2+2x^2+2y^2)^2}{\mu^2+4y^2}+11\mu^2+24x^2+12y^2]\nonumber\\
&&-\frac{3}{y}\int_{0}^{\infty}d\mu\frac{\mu}
{\sqrt{\mu^2+4x^2}}e^{-\sqrt{\mu^2+4x^2}}(\mu^2+2x^2+10y^2)\nonumber\\
&&\times(\mu^2+2x^2+2y^2)
\arctan{\frac{\mu}{2y}}\Bigg]\ ,  \label{eq:N2LO2dr4}\\
v^{2\pi,\rm N2LO}_{\sigma} (r;2\Delta)&=&-\frac{1}
{162 \pi^3  r^6}\frac{(b_3+b_8)\,h_A^3\,y}{(2f_{\pi})^4}\Bigg[
-6\int_{0}^{\infty}d\mu\frac{\mu^2}{\sqrt{\mu^2+4x^2}}e^{-\sqrt{\mu^2+4x^2}}
\nonumber\\
&&\times(\mu^2+4x^2)+\frac{1}{y}\int_{0}^{\infty}d\mu\frac{\mu}
{\sqrt{\mu^2+4x^2}}e^{-\sqrt{\mu^2+4x^2}}(\mu^2+4x^2)\nonumber\\
&&\times(\mu^2+12y^2)
\arctan{\frac{\mu}{2y}}\Bigg]\ , \label{eq:N2LO2dr5}
\end{eqnarray}
\begin{eqnarray}
v^{2\pi,\rm N2LO}_{t\tau} (r;2\Delta)&=&\frac{1}
{1944 \pi^3  r^6}\frac{(b_3+b_8)\,h_A^3\,y}{(2f_{\pi})^4}\Bigg[
-6\int_{0}^{\infty}d\mu\frac{\mu^2}{\sqrt{\mu^2+4x^2}}e^{-\sqrt{\mu^2+4x^2}}\nonumber\\
&&\times(3+3\sqrt{\mu^2+4x^2}+\mu^2+4x^2)+\frac{1}{y}\int_{0}^{\infty}d\mu\frac{\mu}
{\sqrt{\mu^2+4x^2}}e^{-\sqrt{\mu^2+4x^2}}\nonumber\\
&&\times(3+3\sqrt{\mu^2+4x^2}+\mu^2+4x^2)(\mu^2+12y^2)
\arctan{\frac{\mu}{2y}}\Bigg]\ . \label{eq:N2LO2dr6}\\
\end{eqnarray}
\subsection*{Contact terms}
The contact expressions introduced in Section 3.1.2 are defined by the functions $v^{l}_{\rm S}(r)$. These functions are given by

\begin{eqnarray}
v^c_{S}(r)&=& C_S\,f_{\rm short}^{\Delta}(r)+C_1\left[- f_{\rm short}^{\Delta,(2)}(r)-\frac{2}{r}\, f_{\rm short}^{\Delta,(1)}(r)\right]+D_1\left[f_{\rm short}^{\Delta,(4)}(r)+\frac{4}{r}\,f_{\rm short}^{\Delta,(3)}(r)\right]\ , \nonumber\\
\\
v^\tau_{S}(r)&=& C_2\left[- f_{\rm short}^{\Delta,(2)}(r)-\frac{2}{r}\, f_{\rm short}^{\Delta,(1)}(r)\right]+
D_2\left[f_{\rm short}^{\Delta,(4)}(r)+\frac{4}{r}\,f_{\rm short}^{\Delta,(3)}(r)\right]\ , \\
v^\sigma_{S}(r)&=& C_T\,f_{\rm short}^{\Delta}(r)+C_3\left[- f_{\rm short}^{\Delta,(2)}(r)-\frac{2}{r}\, f_{\rm short}^{\Delta,(1)}(r)\right]+
D_3\left[f_{\rm short}^{\Delta,(4)}(r)+\frac{4}{r}\,f_{\rm short}^{\Delta,(3)}(r)\right]\ , \nonumber\\
\\
v^{\sigma\tau}_{S}(r)&=& C_4\left[- f_{\rm short}^{\Delta,(2)}(r)-\frac{2}{r}\, f_{\rm short}^{\Delta,(1)}(r)\right]+
D_4\left[f_{\rm short}^{\Delta,(4)}(r)+\frac{4}{r}\,f_{\rm short}^{\Delta,(3)}(r)\right]\ , \\
v^{t}_{S}(r)&=& -C_5\left[f_{\rm short}^{\Delta,(2)}(r)-\frac{1}{r}\, f_{\rm short}^{\Delta,(1)}(r)\right]+D_5\Bigg[
f_{\rm short}^{\Delta,(4)}(r)+\frac{1}{r}f_{\rm short}^{\Delta,(3)}(r)-\frac{6}{r^2}f_{\rm short}^{\Delta,(2)}(r)\nonumber\\
&&+\frac{6}{r^3}f_{\rm short}^{\Delta,(1)}(r)\Bigg]\ , \\
v^{t\tau}_{S}(r)&=& -C_6\left[f_{\rm short}^{\Delta,(2)}(r)-\frac{1}{r}\, f_{\rm short}^{\Delta,(1)}(r)\right]+D_6\Bigg[
f_{\rm short}^{\Delta,(4)}(r)+\frac{1}{r}f_{\rm short}^{\Delta,(3)}(r)-\frac{6}{r^2}f_{\rm short}^{\Delta,(2)}(r)\nonumber\\
&&+\frac{6}{r^3}
f_{\rm short}^{\Delta,(1)}(r)\Bigg]\ , \\
v^{b}_{S}(r)&=& -C_7\frac{1}{r}f_{\rm short}^{\Delta,(1)}(r)+D_7\left[\frac{1}{r}f_{\rm short}^{\Delta,(3)}(r)+2\,\frac{1}{r^2}f_{\rm short}^{\Delta,(2)}(r)-\frac{2}{r^3}f_{\rm short}^{\Delta,(1)}(r)\right]\ , \\
v^{b\tau}_{S}(r)&=& D_8\left[\frac{1}{r}f_{\rm short}^{\Delta,(3)}(r)+2\,\frac{1}{r^2}f_{\rm short}^{\Delta,(2)}(r)-\frac{2}{r^3}f_{\rm short}^{\Delta,(1)}(r)\right]\ , \\
v^{bb}_{S}(r)&=&-D_9\frac{1}{r^2}
\left[ f_{\rm short}^{\Delta,(2)}(r)-\frac{1}{r}\, f_{\rm short}^{\Delta,(1)}(r)\right] \ ,\\
v^{q}_{S}(r)&=&-D_{10}\frac{1}{r^2}
\left[ f_{\rm short}^{\Delta,(2)}(r)-\frac{1}{r}\, f_{\rm short}^{\Delta,(1)}(r)\right] \ ,
\end{eqnarray}
\begin{eqnarray}
v^{q\sigma}_{S}(r)&=& -D_{11}\frac{1}{r^2}
\left[ f_{\rm short}^{\Delta,(2)}(r)-\frac{1}{r}\, f_{\rm short}^{\Delta,(1)}(r)\right]\ , \\
v^{p}_{S}(r)&=& D_{12}\left[- f_{\rm short}^{\Delta,(2)}(r)-\frac{2}{r}\, f_{\rm short}^{\Delta,(1)}(r)\right]\ , \\
v^{p\sigma}_{S}(r)&=& D_{13}\left[- f_{\rm short}^{\Delta,(2)}(r)-\frac{2}{r}\, f_{\rm short}^{\Delta,(1)}(r)\right]\ , \\
v^{pt}_{S}(r)&=&-D_{14}\left[f_{\rm short}^{\Delta,(2)}(r)-\frac{1}{r}\, f_{\rm short}^{\Delta,(1)}(r)\right]\ , \\
v^{pt\tau}_{S}(r)&=&-D_{15}\left[f_{\rm short}^{\Delta,(2)}(r)-\frac{1}{r}\, f_{\rm short}^{\Delta,(1)}(r)\right] \ ,
\end{eqnarray}
\begin{eqnarray}
v^{T}_{S}(r)&=& C_0^{\rm IT}\,f_{\rm short}^{\Delta}(r)+C_1^{\rm IT}\left[- f_{\rm short}^{\Delta,(2)}(r)-\frac{2}{r}\, f_{\rm short}^{\Delta,(1)}(r)\right]\,\ , \\
v^{\tau z}_{S}(r)&=& C_0^{\rm IV}\,f_{\rm short}^{\Delta}(r)+C_1^{\rm IV}\left[- f_{\rm short}^{\Delta,(2)}(r)-\frac{2}{r}\, f_{\rm short}^{\Delta,(1)}(r)\right]\ , \\
v^{\sigma T}_{S}(r)&=& C_2^{\rm IT}\left[- f_{\rm short}^{\Delta,(2)}(r)-\frac{2}{r}\, f_{\rm short}^{\Delta,(1)}(r)\right]\ , \\
v^{t T}_{S}(r)&=&-C_3^{\rm IT}\left[f_{\rm short}^{\Delta,(2)}(r)-\frac{1}{r}\, f_{\rm short}^{\Delta,(1)}(r)\right] \ , \\
v^{b T}_{S}(r)&=& -C_4^{\rm IT}\frac{1}{r}f_{\rm short}^{\Delta,(1)}(r)\ ,
\end{eqnarray}
where $f_{\rm short}^{\Delta}(r)$ is defined in Section 3.1.2 and $f_{\rm short}^{\Delta,(n)}(r)$ is given by 
\begin{equation}
f_{\rm short}^{\Delta,(n)}(r) =\frac{d^n f_{\rm short}^{\Delta}(r)}{dr^n} \ .
\end{equation}

%\section*{Supplemental Data}
% \href{http://home.frontiersin.org/about/author-guidelines#SupplementaryMaterial}{Supplementary Material} should be uploaded separately on submission, if there are Supplementary Figures, please include the caption in the same file as the figure. LaTeX Supplementary Material templates can be found in the Frontiers LaTeX folder.

%\section*{Data Availability Statement}
%The datasets [GENERATED/ANALYZED] for this study can be found in the [NAME OF REPOSITORY] [LINK].
% Please see the availability of data guidelines for more information, at https://www.frontiersin.org/about/author-guidelines#AvailabilityofData

%%% If you are submitting a figure with subfigures please combine these into one image file with part labels integrated.
%%% If you don't add the figures in the LaTeX files, please upload them when submitting the article.
%%% Frontiers will add the figures at the end of the provisional pdf automatically
%%% The use of LaTeX coding to draw Diagrams/Figures/Structures should be avoided. They should be external callouts including graphics.

%%%%%%%%%%%%%%%% BIBLIOGRAPHY

%\bibliographystyle{frontiersinSCNS_ENG_HUMS} % for Science, Engineering and Humanities and Social Sciences articles, for Humanities and Social Sciences articles please include page numbers in the in-text citations
\bibliographystyle{frontiersinHLTH_FPHY} % for Health, Physics and Mathematics articles
\bibliography{biblio}

%%% Make sure to upload the bib file along with the tex file and PDF
%%% Please see the test.bib file for some examples of references

\end{document}